\documentclass[preprint2,openbib]{aastex6} % double column, single-spaced document
\usepackage{amsmath}
\usepackage{graphicx}
\usepackage{float}
\usepackage{amsfonts}
\usepackage{amssymb}
\usepackage{caption}
\usepackage{subfigure}
\usepackage{overpic}
\renewcommand{\figurename}{\bf{Figure}}

\newcommand{\li}{\uline{\hspace{0.5em}}}
\newcommand{\tabincell}[2]{\begin{tabular}{@{}#1@{}}#2\end{tabular}}

\shorttitle{Properties of a Small-scale Short-duration Solar eruption With a Driven Shock}
\shortauthors{Ying et al.}

\begin{document}
%\begin{thebibliography}{ABCDEFGHIJK}

\title{Properties of a Small-scale Short-duration Solar Eruption with a Driven Shock}
\author{Beili Ying \altaffilmark{1,2,3}, Li Feng\altaffilmark{1,2,4}, Lei Lu\altaffilmark{1}, Jie Zhang\altaffilmark{5}, Jasmina Magdalenic\altaffilmark{6},
   Yingna Su\altaffilmark{1,2}, Yang Su\altaffilmark{1,2}, Weiqun Gan\altaffilmark{1,2}}
\email{lfeng@pmo.ac.cn}

\altaffiltext{1}{Key Laboratory of Dark Matter and Space Astronomy, Purple Mountain Observatory, Chinese Academy of Sciences,210008 Nanjing, China}%
\altaffiltext{2}{School of Astronomy and Space Science, University of Science and Technology of China, Hefei, Anhui 230026, China}
\altaffiltext{3}{Max-Planck-Institut f$\mathrm{\ddot{u}}$r Sonnensystemforschung, Justus-von-Liebig-Weg 3, D-37077, G$\mathrm{\ddot{o}}$ttingen, Germany}
\altaffiltext{4}{State Key Laboratory of Space Weather, National Space Science Center, Chinese Academy of Sciences, Beijing 100190, China}
\altaffiltext{5}{Department of Physics and Astronomy, George Mason University, Fairfax, VA 22030, USA}
\altaffiltext{6}{Royal Observatory of Belgium, Belgium}

\begin{abstract}
Large-scale solar eruptions have been extensively explored over many years. However, the properties of small-scale events with associated shocks have been rarely investigated. We present the analyses of a small-scale short-duration event originating from a small region. The impulsive phase of the M1.9-class flare lasted only for four minutes. The kinematic evolution of the CME hot channel reveals some exceptional characteristics including a very short duration of the main acceleration phase ($<$ 2 minutes), a rather high maximal acceleration rate ($\sim$50 km s$^{-2}$) and peak velocity ($\sim$1800 km s$^{-1}$). The fast and impulsive kinematics subsequently results in a piston-driven shock related to a metric type II radio burst with a high starting frequency of $\sim$320 MHz of the fundamental band. The type II source is formed at a low height of below $1.1~\mathrm{R_{\odot}}$ less than $\sim2$ minutes after the onset of the main acceleration phase. Through the band split of the type II burst, the shock compression ratio decreases from 2.2 to 1.3, and the magnetic field strength of the shock upstream region decreases from 13 to 0.5 Gauss at heights of 1.1 to 2.3 $~\mathrm{R_{\odot}}$. We find that the CME ($\sim4\times10^{30}\,\mathrm{erg}$) and flare ($\sim1.6\times10^{30}\,\mathrm{erg}$) consume similar amount of magnetic energy. The same conclusion for large-scale eruptions implies that small- and large-scale events possibly share the similar relationship between CMEs and flares. The kinematic particularities of this event are possibly related to the small footpoint-separation distance of the associated magnetic flux rope, as predicted by the Erupting Flux Rope model.

\end{abstract}


\section{Introduction}
\bibliographystyle{apj}

Coronal mass ejections (CMEs) and flares are two most important eruptive phenomena in the atmosphere of the Sun, and are now regarded as the primary solar drivers of geomagnetic storms and ionospheric disturbances on the Earth. Theoretically and observationally, many researches found that a CME may originate as a magnetic flux rope (MFR) which contains a coherent magnetic structure with magnetic field winding around its central axis \citep[e.g.,][]{Chen1989,Zhangjie2012,Wanghm2015}. In the Atmospheric Imaging Assembly \citep[AIA;][]{Lemen2012} field of view (FOV), MFRs can often be observed in 131~{\AA} and/or 94~{\AA} passbands \citep[e.g.,][]{Cheng2013a,Cheng2014}. Using AIA multi-wavelength observations of a flaring active region, \citet{Reeves2011} has constrained the temperature of MFRs to be between 5 MK and 20 MK, which suggests that the MFRs could reach flare-like temperatures. When MFRs are observed edge-on above the limb, they show up as dark cavities characterized by low densities and high temperatures \citep{Gibson2006,Gibson2010,Kucera2012}, exhibiting signatures of spiral motion of plasma \citep[see details in][]{Schmit2009}. Due to the property of its high temperature, the MFR observed in the AIA images is also termed as a ``hot channel (HC)'' \citep{Zhangjie2012}. When the HC transits into the FOV of the Large Angle Spectroscopic Coronagraph \citep[LASCO;][]{ Brueckner1995}, it often appears as a three-part CME structure \citep{ Illing1983}.

An MFR can form either before or during an eruption. Many researches have found indirect evidences of the existence of MFR before the eruption, e.g., forward or reverse S-shaped sigmoids \citep{Rust1996,Canfield1999,Tripathi2009}, dark cavities \citep{Low1995,Gibson2004,Dove2011}, as well as filaments \citep{Mackay2010,Guo2010,Su2011}. \citet{Patsourakos2013} presented the first direct evidence of a fast CME driven by the destabilization of a preformed coronal MFR. \citet{Song2014} witnessed the formation of an MFR during the eruption.
Theoretically, CMEs have been presumed as the eruption of MFRs: some began with initial coronal FRs in equilibrium \citep{Chen1989,Vrsnak1990,Chen1993,Gibson1998,Roussev2003}, while others started with the emergence of FRs from below the photosphere \citep{Wu1997,Fan2003,Manchester2004}.
Among the existing CME models, the semi-analytic erupting flux-rope (EFR) model of CMEs \citep{Chen1989,Chen2010} has been tested widely. The basic physical principle of the EFR model is that the major radial Lorentz self-force (the Lorentz ``hoop force'') accelerates coronal FRs. The EFR model introduced the distance $S_f$ between the two footpoints of the MFR at the base of the corona. It is suggested that $S_f$ is a key parameter to determine the dynamics of MFRs.

When the speed of a CME is higher than the local fast magnetosonic speed, a shock wave may be driven. If the CME as a driver pushes the plasma and its body expands in all directions; meanwhile, the offset distance between shock front and CME ejecta always increases in time, this type of shock can be regarded as a three-dimensional piston-driven shock \citep{Vrsnak2008}. Shock waves are accompanied by a type II radio burst observed in the radio dynamic spectra as emission lanes slowly drifting to lower frequencies with time. The type II bursts classically consist of two bands separated by approximately a factor of 2 in frequency. The lower frequency (fundamental) band is approximately at the local plasma frequency $f_p$, while the upper (second harmonic) band is roughly at 2$f_p$. The starting frequency of the metric type II emission associated with the coronal shock is usually less than 300 MHz with an average of about 100 MHz \citep{Gopalswamy2005}. High starting frequency implies that the shock is propagating through a high density region and hence a low formation height. Some of the coronal shock waves propagate in the interplanetary space and become the source of type II radio radiation in the decametric to kilometric wave range \citep{Gopalswamy2005,Lu2017,Prakash2017}. In some of the type II burst events, fundamental and/or harmonic bands are split into two morphologically similar lanes \citep{Nelson1985}. A number of explanations were suggested accounting for the magnetic, Doppler, and geometrical effects \citep{Krueger1979}. \citet{Smerd1974} attributed the band split to the emission from the upstream and downstream shock regions, demonstrating a potentially vital method that could offer an estimation of the corona Alfv\'{e}n velocity and magnetic field strength.

Mini-filaments are small-scale features in the solar atmosphere, which frequently occur across the entire disk. They have a spatial scale of a few tens of arcsecs and are the small-scale analog to large-scale filaments \citep{Wang2000}. Mini-filaments located at the edge of active regions often act as triggers of coronal jets \citep{Hong2016,Wyper2017}. It is rare that these mini-filaments are observed to be associated with CME HCs. In this paper we present the analyses of such a CME event and its driven shock. Recently, \citet{Kumari2017} analyzed the same CME-driven shock, and used the shock to represent the near-Sun kinematics of the CME. In our study, we have not only investigated the shock properties much more extensively, but also give the interpretations on what causes the particularities of this eruption event. We have quantified the characteristics of various eruption features: CME HC, its driven bright front as the likely location of the shock, and the associated flare. As the distance of footpoint separation of an MFR is suggested to be a key parameter to determine its dynamics, we will investigate how the analyzed eruption originating from a small source region differs from large-scale eruptions. Previous studies have found equal partition of free magnetic energies between flares and CMEs in large-scale eruption events \citep{ Emslie2012,Feng2013}. Similar energy estimate will also be carried out to check whether our small-scale eruption follows the same rule.

The organization of the paper is as follows. In Section 2, we describe the observations of our small-scale eruptive events taken by different instruments. In Section 3, the properties of the HC and associated bright front/shock front, the impulsive compact flare, the CME travelling into LASCO FOV are analyzed. The deduced coronal magnetic field  is included as well. The last section summarizes the results and discusses the differences and similarities between small-scale and large-scale eruptions.

%\captionsetup{font=footnotesize,justification=raggedright,labelsep=period}
%\renewcommand{\captionlabeldelim}{.}

\section{Instruments and Observations}
\subsection{Instruments}
\begin{figure*}[htb]
\centering
  \includegraphics[width=10. cm, angle=90]{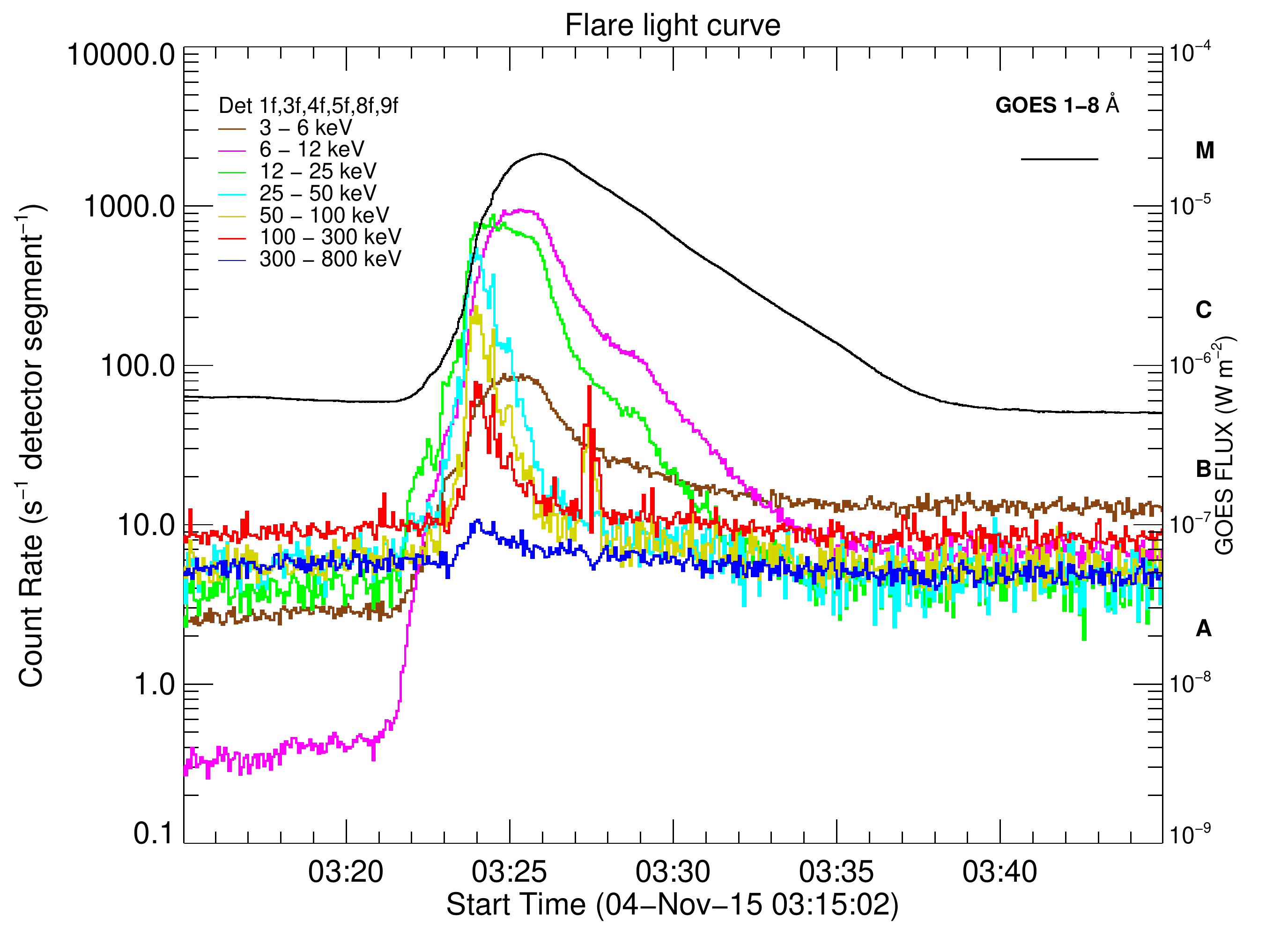}\\
 \caption{The light curves of GOES 1-8~{\AA} and RHESSI in multiple energy bands up to 800 keV. }
 \label{fig:RHE}
\end{figure*}
The Atmospheric Imaging Assembly (AIA) launched as a part of NASA's \textit{Solar Dynamics Observatory (SDO)} mission provides high temporal (12 seconds) and spatial resolution (0.6" per pixel) full-disk images of the corona at multiple wavelengths covering the temperature from $\sim5000$ K to $\sim20$ MK. The vector magnetogram data from the Helioseismic and Magnetic Imager (HMI) \citep{Scherrer2012} aboard the SDO reveals the magnetic field context of the source region of the eruption event. To examine the Soft X-ray (SXR) emission of the associated flare, we study the data from \textit{Geostationary Operational Environmental Satellite (GOES)}, and discuss the Hard X-ray (HXR) emission through the \textit{Reuven Ramaty High-Energy Solar Spectroscopic Imager (RHESSI)} mission \citep{Lin2002}. For the associated CME, we use the data from the Large Angle Spectroscopic Coronagraph (LASCO) C2 aboard the \textit{Solar and Heliospheric Observatory(SOHO)} with the field of view (FOV) 1.5-6 $R_{\odot}$. In order to investigate the type II radio burst associated with the CME-driven shock, we utilize the radio dynamic spectra from Learmonth Observatory in Australia, with the frequency range of 25-180 MHz, and the temporal resolution of 3 seconds, and from Culgoora observatory whose frequency range of 18-1800 MHz.
\begin{figure*}[htb]
\centering
  \includegraphics[width=1.\textwidth]{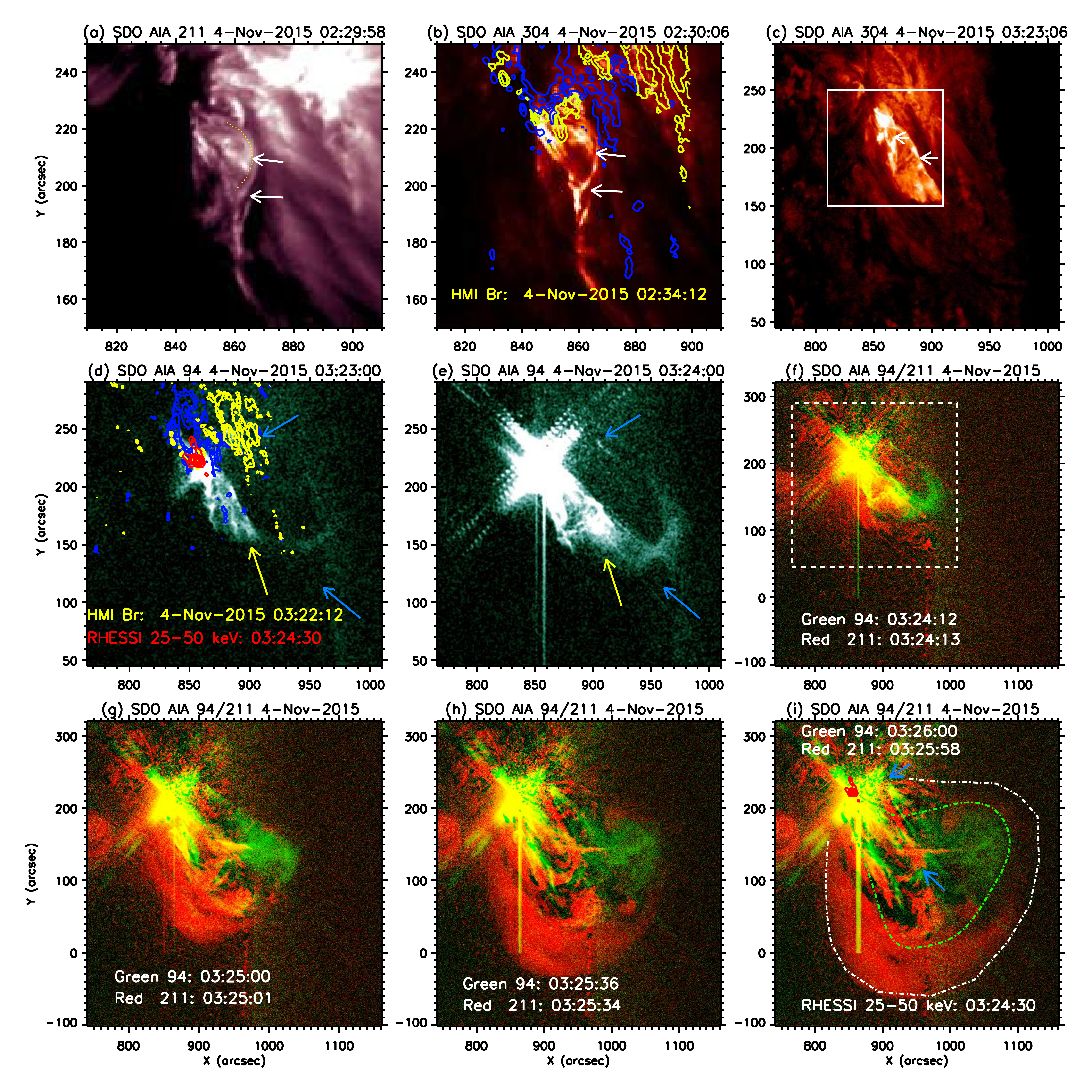}\\
 \caption{(a)-(c) Evolution of the mini-filaments as observed in AIA 211 {\AA} ($\sim$2 MK) and 304 {\AA} ($\sim$0.05 MK). The two white arrows mark the positions of two mini-filaments or their segments. The white box in panel (c) defines the region of the interest (ROI) enlarged in panels (a) and (b). Panel (c) is an AIA 304 {\AA} image observed at~03:23 UT at the beginning of the eruption, while panels (a) and (b) are AIA images observed in 304 {\AA} and 211 {\AA}, respectively, at 02:30 UT before the eruption. In panel (a), one of the mini-filaments is delineated by a yellow dotted line. In panel (b), the blue (positive) and yellow contours (negative) show HMI radial magnetic field at 02:34 UT. (d)-(e) AIA 94 {\AA} ($\sim$7 MK) base difference images displaying the eruption of the HC at 03:23:00 UT and 03:24:00 UT. Two blue arrows point to the same coordinates in panels (d), (e), (i). The upper arrow indicates a fixed footpoint of a coronal loop heated by the accelerated particles. The lower arrow shows the top evolution of the loop. The white box in panel (f) defines the ROI enlarged in panel (c), (d) and (e). SDO/HMI radial magnetic field (blue: positive and yellow: negative) is at 03:22:12 UT in panel (d). The contour levels of the RHESSI 25-50 keV source (red contours) are 30\%, 40\%, 50\%, 70\%, 90\% in panels (d) and (i). The RHESSI source is at 03:24:30 UT around flare peak time. Yellow arrows represent the HC front in panels (d) and (e). (f)-(i) The synthesized base difference images from~03:24 UT to~03:26 UT with the 211 {\AA} image in the red channel and the 94 {\AA} image in the green channel. In panel (i), a green dotted-dashed profile indicates the periphery of the HC at 03:26:00 UT, and a white curve shows the periphery of the EUV bright front at 03:25:58 UT. For all the base difference images, a pre-event image at 03:20:00 UT is subtracted.}
 \label{fig:HC}
\end{figure*}

\begin{figure*}[htb]
  \centering

  %\subfigure{
  \begin{minipage}[b]{0.49\linewidth}
  \raggedleft
  \begin{overpic}[width=8. cm]{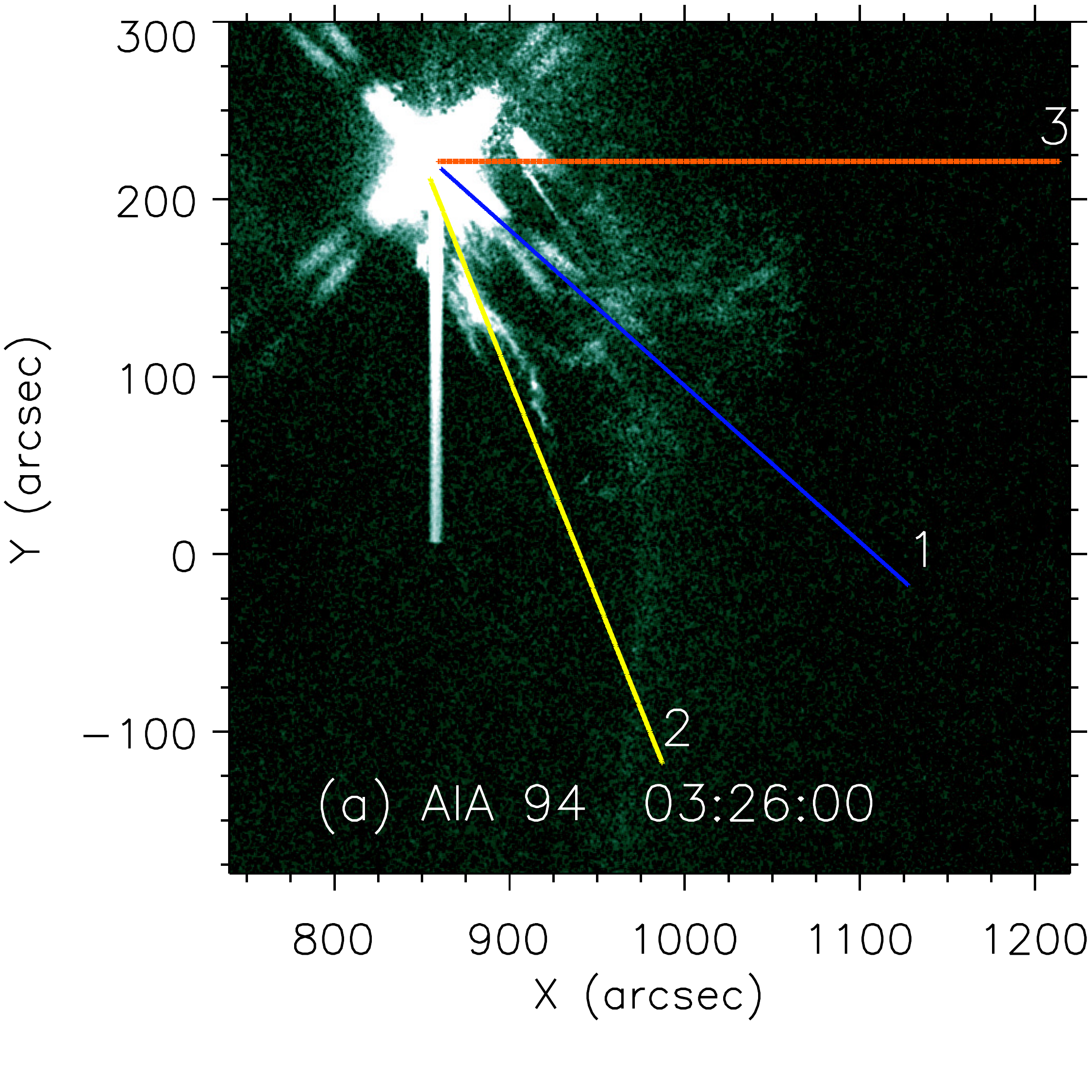}
     %\put(85,88){\color{white}{(a)}}
     %\put(42,30){\large\color{yellow}$\Rightarrow$}
     \end{overpic}
    \end{minipage}
  %\centering
  \begin{minipage}[b]{0.49\linewidth}
  \raggedright
    \begin{overpic}[bb=70 1 715 715,clip,width=8. cm]{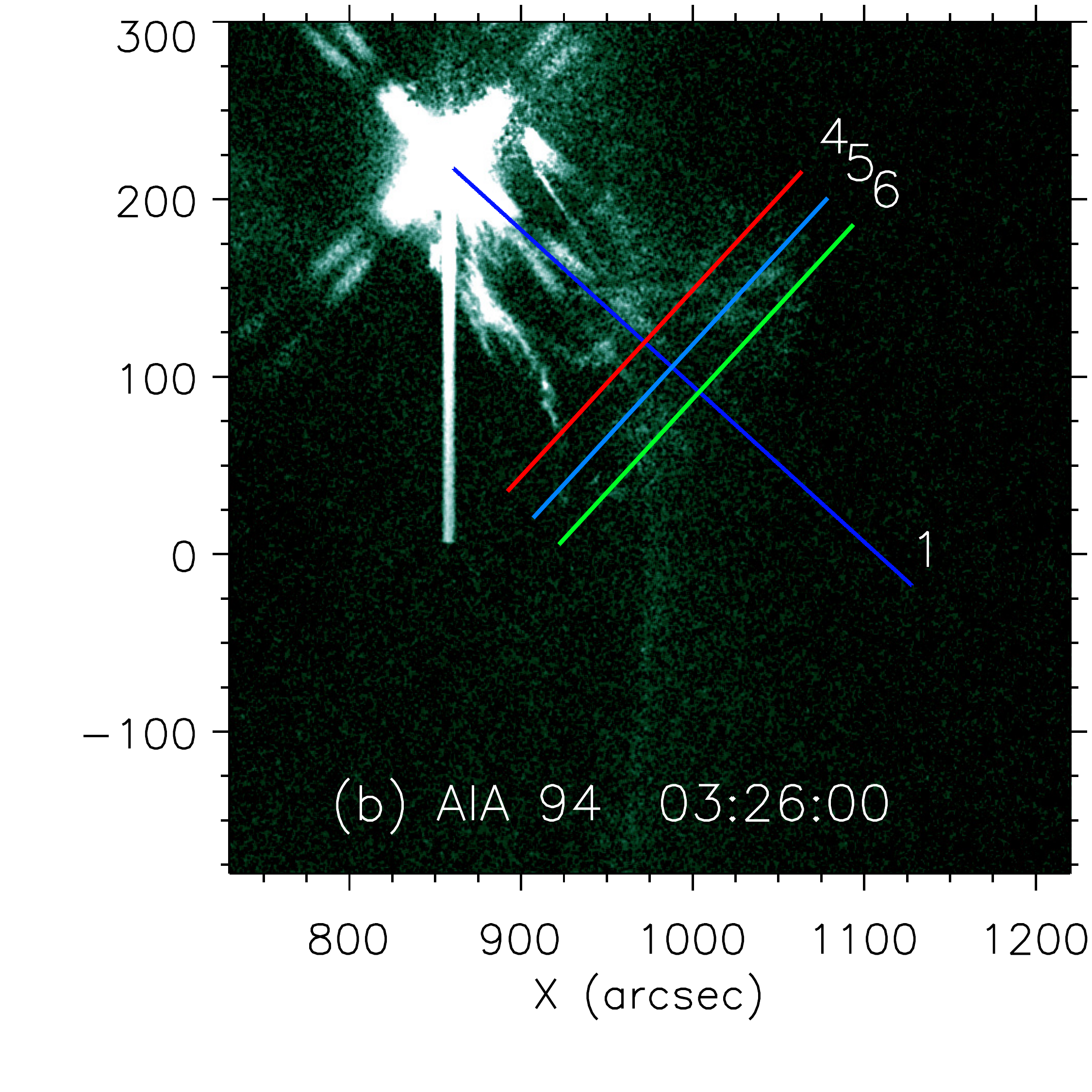}
     %\put(85,88){\color{white}{(b)}}
     %\put(42,30){\large\color{yellow}$\Rightarrow$}
     \end{overpic}
    \end{minipage}
%  }

 \caption{SDO/AIA 94 {\AA} ($\sim$7 MK) base difference images displaying the eruption of the HC at 03:26:00 UT. In panel (a), solid lines 1, 2 mark two selected slices along the propagation direction of the HC nose and its flank. Slice 3 is chosen to track the HC deflection. The other lines 4-6 represent three slices along the expansion directions of the HC flank in panel (b). The width of slice 3 is 7 arcseconds and that of other slices is 2 arcseconds.}
 \label{fig:shock}
\end{figure*}

\subsection{Event Overview}

On November 4, 2015, an eruption occurred from a small region at the edge of NOAA active region 12445 (W70$^\text{o}$, N15$^\text{o}$) located close to the west solar limb. The GOES 1-8~{\AA} SXR flux indicated by the black line in \autoref{fig:RHE} shows that the M1.9-class flare in this event has a short duration of about 18 minutes, and its flux starts to rise at~03:21 UT and peaks at 03:26~UT. RHESSI photon light curves in different energy bands within 3-800~keV are also presented in \autoref{fig:RHE}. The enhancement of the photon count rates in the high energy band of 300-800~keV implies that the electrons were probably accelerated to high energies.
\begin{figure*}[htb]
\noindent
\centering
\hspace{1. cm}
\subfigure{
 %\centering
  \begin{minipage}[b]{0.45\linewidth}

   \begin{overpic}[bb=5 65 640 780,width=1.02\textwidth]{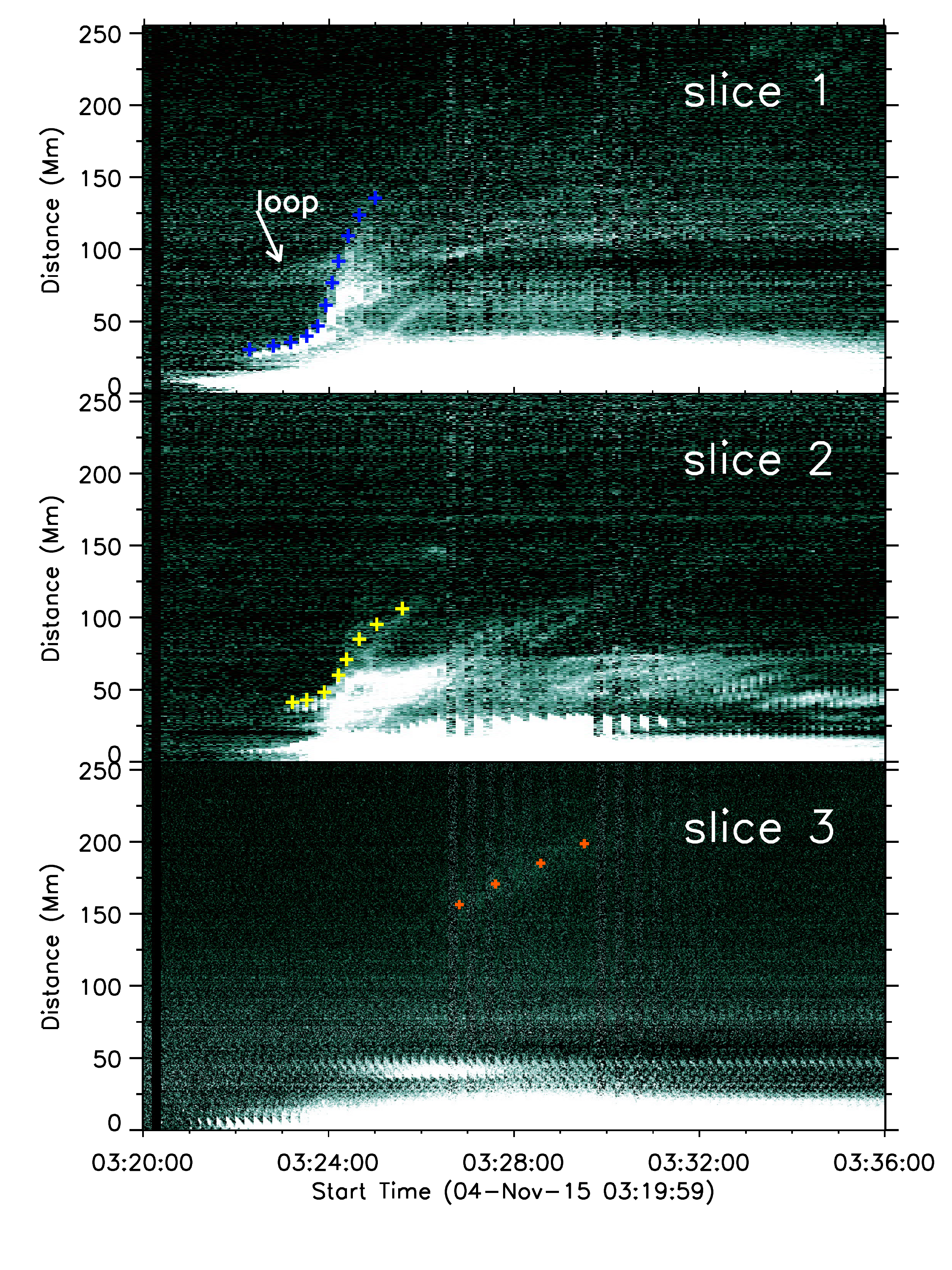}
   \put(72,14){\color{white}{(a)}}
   %\put(40,14){\color{white}{\Downarrow}}
   %\put(82,14){\color{white}{\Downarrow}}
   \end{overpic}

  \end{minipage}
  %\centering
  \begin{minipage}[b]{0.45\linewidth}

   \begin{overpic}[width=1.0\textwidth]{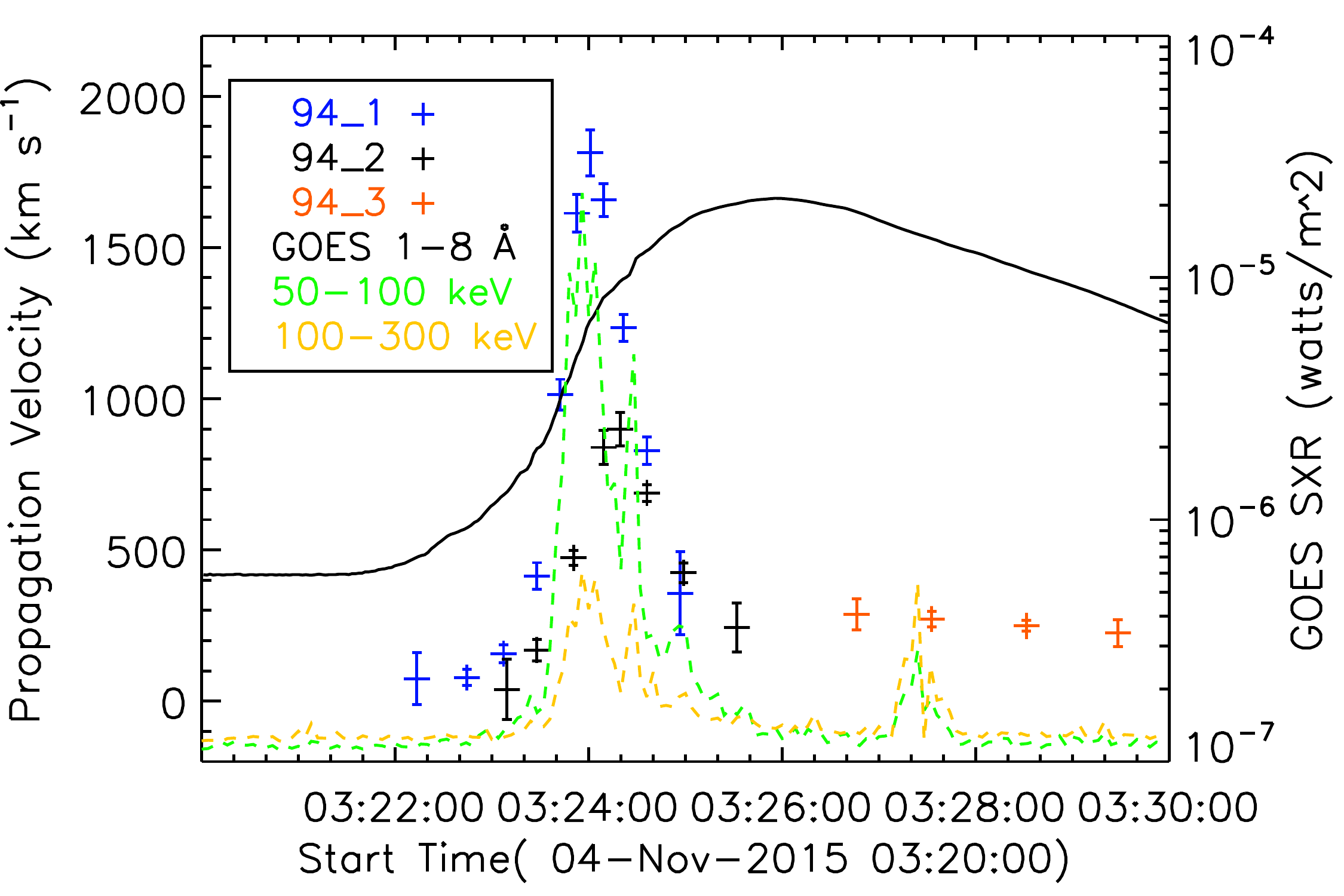}
   \put(75,15){(b)}
   \end{overpic}

    \begin{overpic}[width=.94\textwidth]{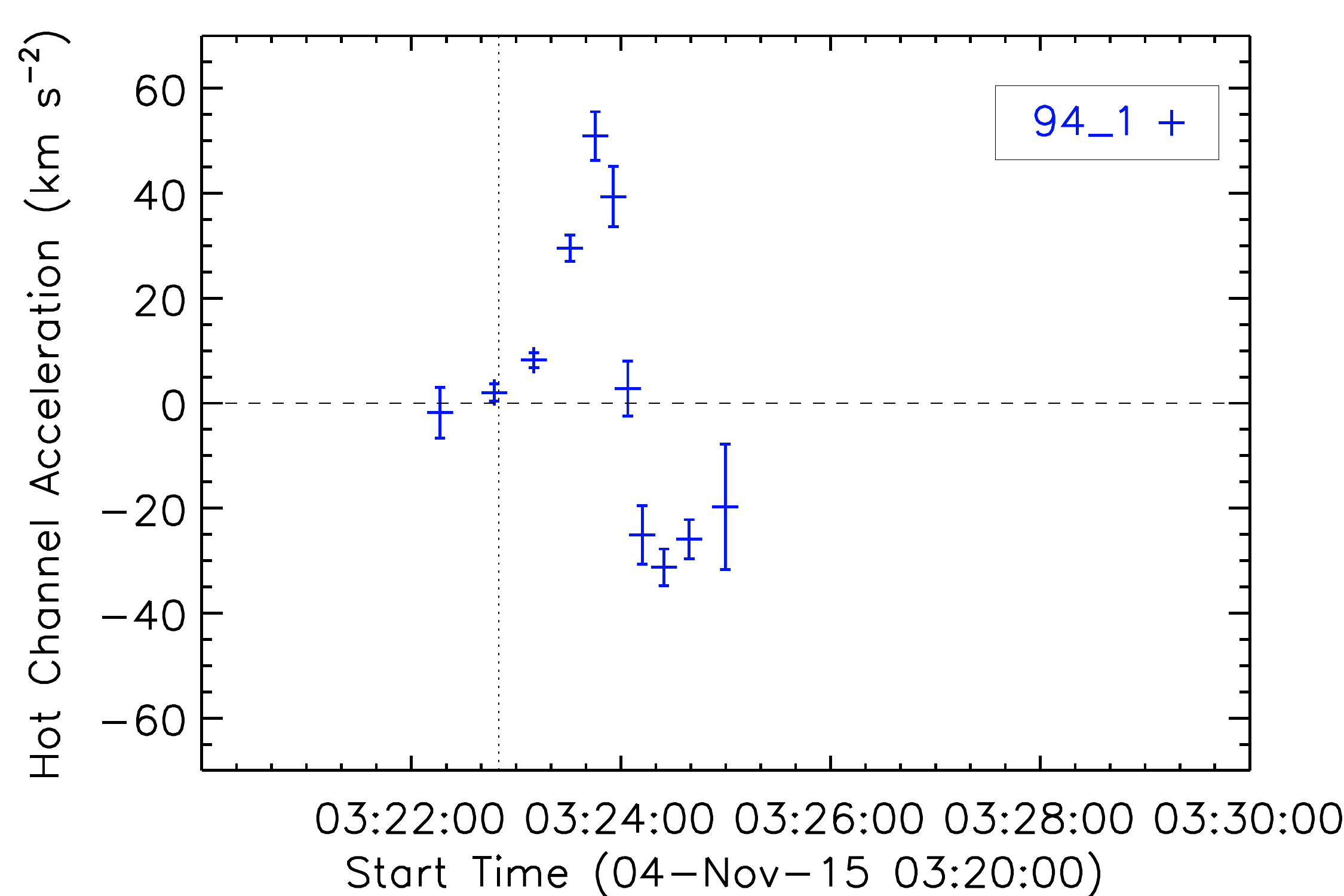}
     \put(82,15){(c)}
    \end{overpic}
  \end{minipage}%
  }

\caption{(a) Base difference distance-time plots of the 94 {\AA} passband along the slices 1, 2 and 3 as shown in \autoref{fig:shock} (a). The track marked by the white arrow denotes the distance evolution of the coronal loop. (b) Velocity evolution of the HC along slices 1, 2, and 3. The black-solid profile denotes the GOSE 1-8 {\AA} SXR flux of the associated flare, and the yellow (green) curve is the photon count rates in the high energy band of 100-300 (50-100) keV. (c) The evolution of acceleration along slice 1. The vertical dotted lines indicate the transit time at $\sim$03:22:50 UT between the slow rise and fast rise phases. The error bars along slices 1 and 2 (slice 3) in panels (b) and (c) are propagated from the uncertainty of $\pm\,6  (12)\,\mathrm{Mm}$ in the distance measurements.}
\label{fig:SHC_vel}
\end{figure*}

\autoref{fig:HC} is an overview of the AIA observations of the event including the evolution of the mini-filaments (a-c), CME HC (d-i), and a bright front preceding the HC (f-i). Panels (a-c) present the evolution of mini-filaments in AIA 211 {\AA} ($\sim$2 MK) and 304 {\AA} ($\sim$0.05 MK) before the eruption at $\sim$02:30 UT and in 304 {\AA} $\sim$03:23 UT during the eruption. The white box in panel (c) defines the region of the interest (ROI) enlarged in panels (a) and (b). Two white arrows mark the positions of two mini-filaments or their segments. Actually quite a few mini-filaments were observed in this event. As an example, in panel (a) one of the mini-filaments is delineated by a yellow dotted line. In panel (b), blue and yellow contours indicate the positive and negative radial magnetic fields of AR 12445 observed by HMI at 02:34:12 UT. As it is rather close to the limb, to reduce projection effects, the radial component derived from the HMI vector magnetogram instead of the line-of-sight magnetic field is plotted.  As revealed by panel (c), the mini-filaments in the white box begins to evolve slowly at $\sim$03:23 UT. We conjecture that these mini-filaments may be related to the eruption process.

\autoref{fig:HC} (d)-(e) are AIA base-difference images in 94 {\AA} ($\sim$7 MK) at 03:23:00 UT and 03:24:00 UT produced by subtracting a pre-event image at 03:20:00 UT. In \autoref{fig:HC} (d), the colors of HMI radial magnetic field contours have the same definition as in \autoref{fig:HC} (b). The location of the flaring region is consistent with that of the RHESSI HXR source in 25-50 keV, marked by the red contours with levels of 30\%, 40\%, 50\%, 70\%, 90\% of the maximal intensity of RHESSI image. The CLEAN image reconstruction method \citep{Hurford2002} is applied to produce the RHESSI source, and the selected time interval for imaging is close to from 03:24:30 to 03:25:30 UT, which is close to the HXR peak time. Combining observations of magnetic field, EUV, SXR and HXR images, we can see that the source region of the eruption is located in a region of about $25\times50$ arcsec$^2$ with mixed magnetic polarities. Blue arrows in \autoref{fig:HC} (d), (e) and (i) have the same coordinates. The upper arrow indicates the fixed footpoint of a coronal loop. As time elapses, the footpoint becomes brighter and brighter. We infer that the accelerated electrons were released from the flaring region, transported along the magnetic field lines indicated by the coronal loop structure, and deposited energy at the other footpoint. The lower arrow is to mark the near top of the loop, and the change of the stand-off distance between the loop top and the arrow shows the height evolution of the loop top. Yellow arrows in \autoref{fig:HC} (d) and (e) indicate the position of the HC front.

\autoref{fig:HC} (f-i) are synthetic images with AIA 94 {\AA} base-difference images in green channel and AIA 211 {\AA} base-difference images in red channel observed from 03:24 UT to 03:26 UT. \autoref{fig:HC} (i) displays the synthesized image at the GOES peak time. Both a bubble-like HC bounded by a green curve and a bright front preceding the HC bounded by a white curve can be clearly seen. Following the evolution from \autoref{fig:HC} (f) to (i), we can observe the northward deflection of HC  which is further validated by the LASCO image in Section 3.4.

\begin{figure*}[htb]
%\centering
 \subfigure{
 %\centering
  \begin{minipage}[b]{0.5\linewidth}
  \raggedleft
   \begin{overpic}[width=0.8\textwidth]{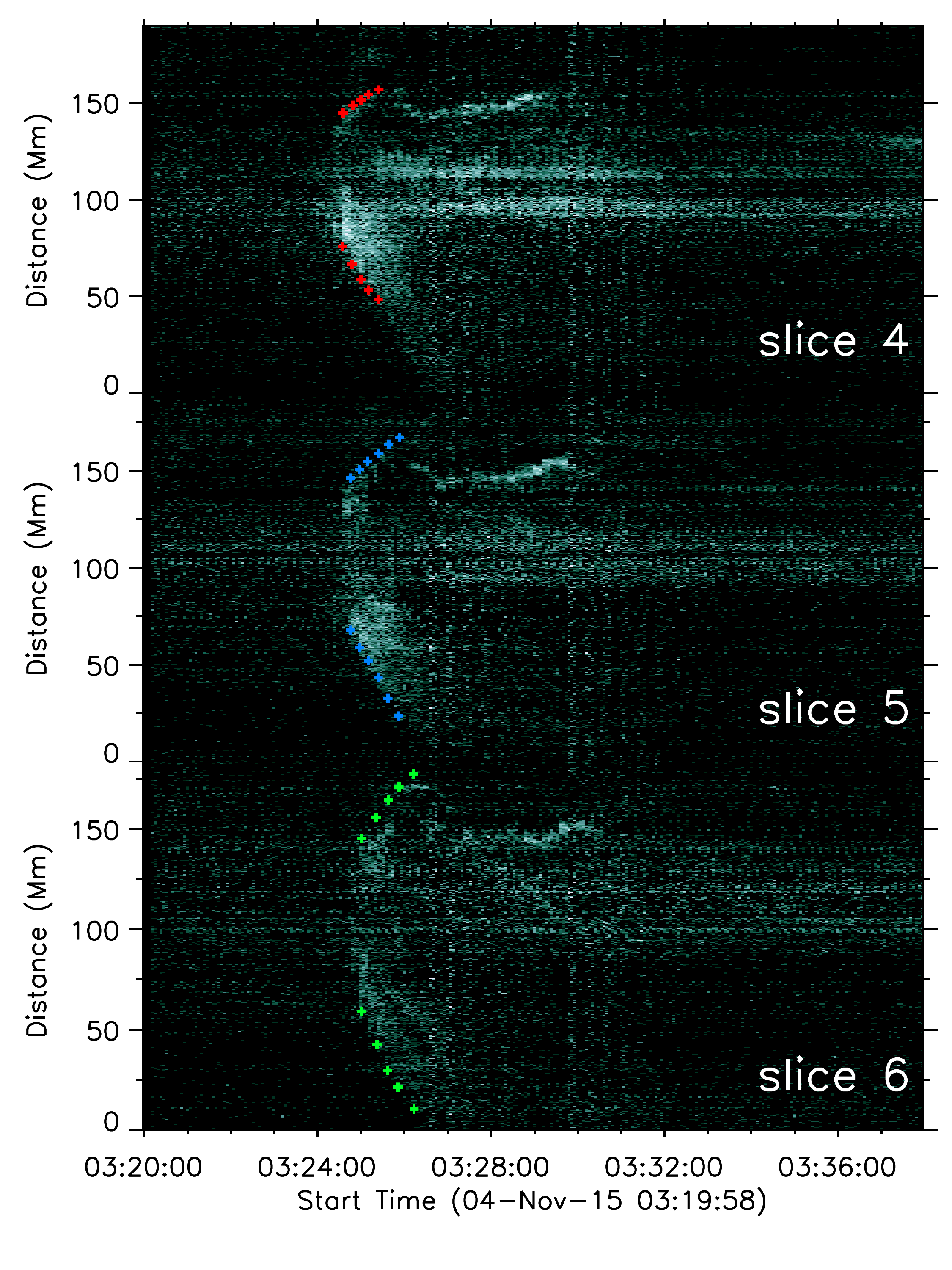}
   \put(62,93){\color{white}{(a)}}
   \end{overpic}
  \end{minipage}
  %\centering
   \begin{minipage}[b]{0.5\linewidth}
   \raggedright
   \begin{overpic}[width=0.8\textwidth]{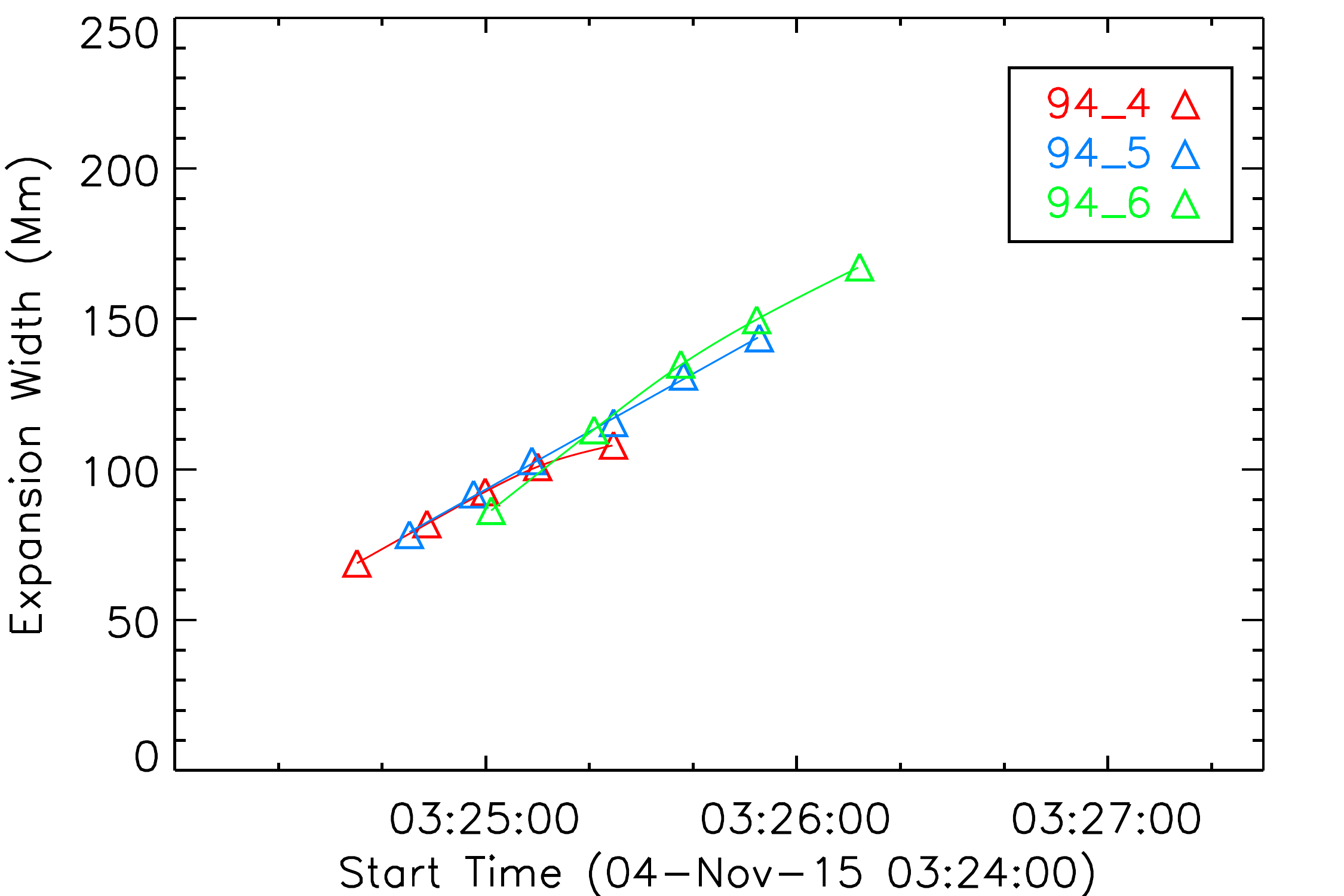}
   \put(81,12){(b)}
   \end{overpic}
   \begin{overpic}[width=0.8\textwidth]{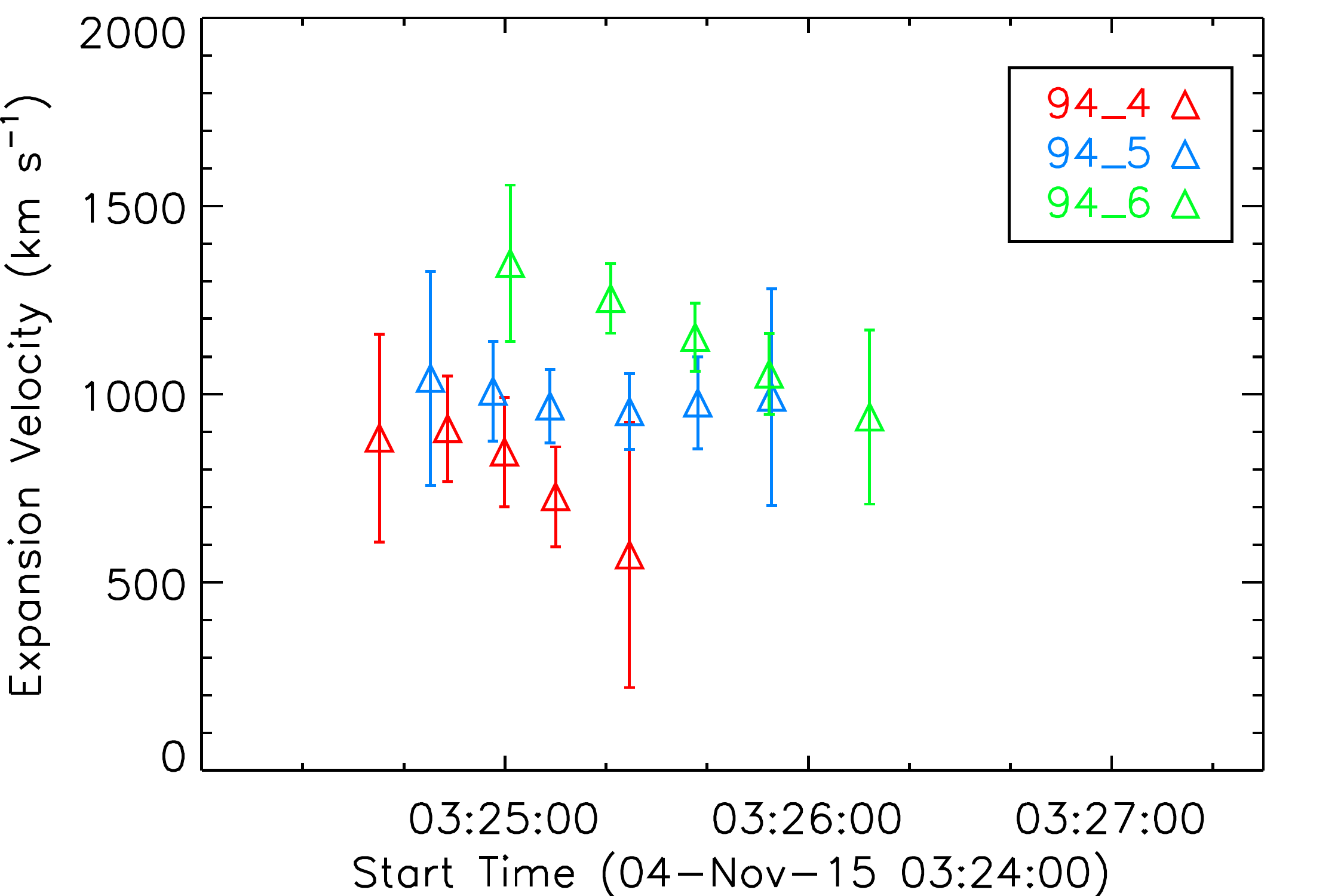}
   \put(82,12){(c)}
   \end{overpic}
  \end{minipage}

 }
\caption{(a) Base difference distance-time plots of the 94 {\AA} passband along slices 4-6 as shown in \autoref{fig:shock} (b). (b) The evolution of the HC width along slices 4-6 with different colors obtained from the data in panel (a). (c) The evolution of the HC expansion velocity derived from the data in panel (b). The error bars in panel (c) are propagated from the uncertainty of $\pm\,6\,\mathrm{Mm}$ in the distance measurements.}
\label{fig:HC_exp_vel}
\end{figure*}

\section{Analysis and Results}

\subsection{Propagation and Expansion of the Hot Channel}
%LI Feng
In order to display the rising motion of the HC nose and flank as well as the lateral expansion of the HC flank, we select two slices (1 and 2) along its  propagation direction as shown in \autoref{fig:shock} (a), and three slices 4 to 6 along the expansion direction of the HC flank which is perpendicular to the main propagation direction (slice 1) as shown in \autoref{fig:shock} (b). Slice 3 is selected to track the HC deflection. In the minutes after $\sim$03:26:30 UT, the change of the HC's direction takes place along slice 3 in \autoref{fig:shock} (a). The width of the slices is $\sim$2 arcseconds (3 pixels)  except for slice 3 ( of which the width is $\sim$7 arcseconds (12 pixels)). The kinematic evolution of the HC along slices 1 to 3 are shown in \autoref{fig:SHC_vel}. In \autoref{fig:SHC_vel} (a), to create the distance-time plots along slices 1 to 3, we stack the base difference intensity along each slice with time. The position of the HC front along slices 1 to 3 is marked by blue, yellow, and orange plus signs, respectively. The track marked by the white arrow in the upper panel indicates the coronal loop introduced in \autoref{fig:HC}. Although the signal-to-noise ratio along slice 3 is very low, we can still observe the appearance of the signal later after 03:26:30 UT. One of the interpretations may be the possible northward deflection of the HC. Fitting the distance-time data points with smoothing cubic splines, we further derive reliable velocities and accelerations. The deduced velocity and acceleration data points are presented in \autoref{fig:SHC_vel} (b) and (c). Concerning the error analyses of the kinematics, we estimate that the maximal uncertainty in the distance measurement along the track of HC front is about 6 Mm (12 Mm) for slices 1 and 2 (slice 3) by randomly tracing the front for a few times. Thus, we assume the distance uncertainty is 6 Mm (12 Mm) above and below each point along the track for slices 1 and 2 (slice 3). Then this uncertainty is propagated to the uncertainties in velocity and acceleration with 100 Monto Carlo simulations. In such simulations, 100 randomly selected distance values following a Gaussian distribution with $3\sigma = 6$ Mm (12 Mm along slice 3) are used at each distance-time data point. Then the velocity and acceleration are derived 100 times using these randomly distributed distance values with the same procedure described above. The derived error bars are indicated in \autoref{fig:SHC_vel} (b) and (c).

 \autoref{fig:SHC_vel} reveals that the kinematics of the HC along its main propagation direction (slice 1) can be clearly divided into three phases: a slow rise phase before 03:22:50~UT with small velocity and acceleration, a following fast rise phase from 03:22:50~UT to 03:24:10~UT with a peak velocity of about 1800 km  $\mathrm{s}^{-1}$ and a peak acceleration of about 50 km $\mathrm{s}^{-2}$. The second phase is the main acceleration phase. The third phase after 03:24:10~UT is the residual acceleration phase and is characterized by a peak deceleration of about 30 km $\mathrm{s}^{-2}$ (more details can see \autoref{table:phase} ).
 \begin{table*}[!ht]

  \centering
    \caption{Different phases of the HC $\&$ the bright front (BF) along slice1}
     \label{table:phase}
     %\begin{threeparttable}[b]
  %\begin{small}
  \begin{minipage}[t]{0.65 \linewidth}

  \renewcommand{\arraystretch}{1.4}
    \begin{tabular}{p{2.5 cm}<{\centering} p{5. cm}<{\centering} p{3. cm}<{\centering} }
    \hline
     \hline
    \tabincell{c}{Time }                                                   & \tabincell{c}{HC phase\tablenotemark{a} }   & \tabincell{c}{BF phase\tablenotemark{b}} \\
    \hline
    \tabincell{c}{03:22:00}                                              & Commencement                                          & --- \\
    %\hline
    \tabincell{c}{03:22:00-03:22:50 }                              & \tabincell{c}{Slow rise (initiation)}                     & ---  \\
    %\hline
    \tabincell{c}{03:22:50-03:24:10\tablenotemark{c}  } & \tabincell{c}{ Fast rise (main acceleration)} & \tabincell{c}{ Fast rise } \\
     %\hline
    \tabincell{c}{03:24:10\tablenotemark{d}  }                & \tabincell{c}{Peak speed}                            & \tabincell{c}{Peak speed} \\
    %\hline
    \tabincell{c}{03:24:10-03:33:30}                               & \tabincell{c}{Residual acceleration}             & \tabincell{c}{ Deceleration phase}\\
    %\hline
     %\tabincell{c}{03:26:30-03:29:30}                              & \tabincell{c}{Change of direction}                & --- \\
   % \hline
    \tabincell{c}{03:25:00-03:39:00}                               & ---                                                                & \tabincell{c}{ Type II burst lifetime}  \\

    \hline
   \end{tabular}
   %\raggedleft

   \small{Notes:}
   \footnotetext{{The results of the HC are derived from AIA 94 {\AA} images.}}
   \footnotetext{{The results of the BF are derived from AIA multi-band images.}}
    \footnotetext{{The spatial separation time between the HC and the BF along slice 2 is at 03:23:10 UT.}}
    \footnotetext{{The separation time in space between the HC and the BF along slice 1.}}
   %\end{threeparttable}
%\end{sidewaystable}
 % \end{small}
   \end{minipage}

\end{table*}

\autoref{fig:SHC_vel} (b) demonstrates that the HC main acceleration phase coincides very well with the flare impulsive phase as indicated by the RHESSI light curves. It implies that the impulsive acceleration phase of the HC is closely related to the magnetic reconnection process. This scenario is in general consistent with the CME-flare temporal relationship described in \citet{Zhang2001} and \citet{Zhang2006}. However, there are a few particularities we need to point out for this event.
\begin{itemize}
\item[-] The maximal acceleration rate of about 50 km $\mathrm{s}^{-2}$ during the main acceleration phase is two orders of magnitude larger than the average rate of 331 m $\mathrm{s}^{-2}$ given in \citet{Zhang2006} and one order of magnitude larger than their maximal value of 4464 m $\mathrm{s}^{-2}$.

\item[-] The duration of the main acceleration phase is no more than two minutes, which is smaller than the minimal value of 6 minutes in \citet{Zhang2006}.

\item[-] The acceleration magnitude (A) and the duration (T) is in general following the rule proposed by \citet{Zhang2006} that the shorter the duration, the larger the acceleration. However, the magnitude A of 50 km $\mathrm{s}^{-2}$ is higher than the value derived from the scaling law $\mathrm{A}(\mathrm{m}~\mathrm{s}^{-2}) = 10,000 \mathrm{T}^{-1}$(minute) derived in \citet{Zhang2006}.

\item[-] The deceleration of the HC in the residual acceleration phase with maximal value of about 30 km $\mathrm{s}^{-2}$ is much larger than the reported value of  131 m $\mathrm{s}^{-2}$ in \citet{Zhang2006}.
\end{itemize}
%\noindent

\autoref{fig:HC_exp_vel} presents the lateral expansion properties of the HC flank. \autoref{fig:HC_exp_vel} (a) includes three stack plots of the AIA 94~{\AA} base-difference intensity along slices 4-6. The derived HC widths along these three slices are demonstrated in \autoref{fig:HC_exp_vel} (b) in different colors. The corresponding expansion velocities and their associated error bars are calculated with the same method as used for the propagation. \autoref{fig:HC_exp_vel} (c) reveals that the expansion velocity has a peak value of about 1350 km $\mathrm{s}^{-1}$ along slice 6 at 03:25:00 UT, and slows down to about 1000 km $\mathrm{s}^{-1}$.

\begin{figure*}[htb]
\centering
\includegraphics[width=1. \textwidth]{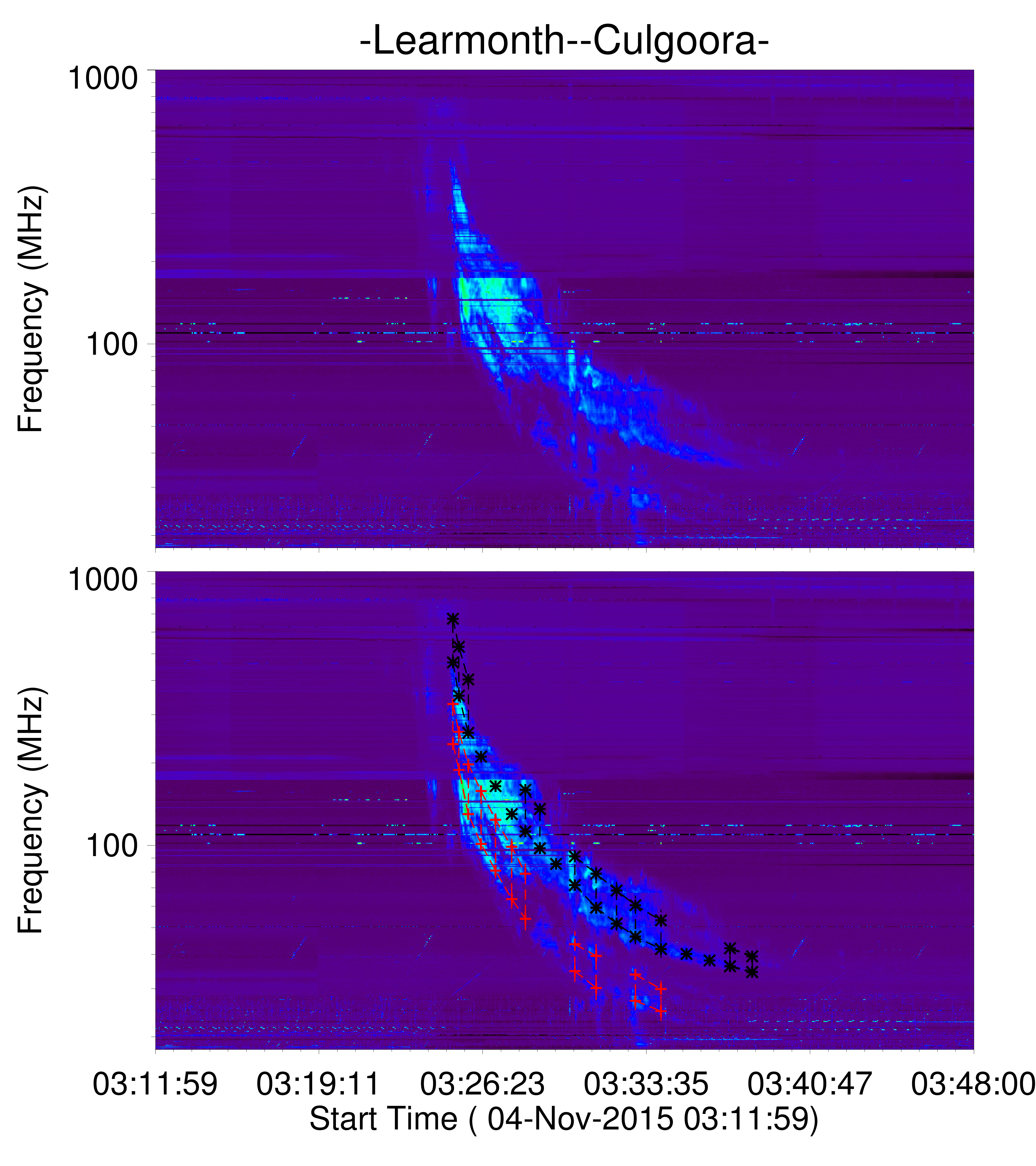}
\caption{Radio dynamic spectrum from the Learmonth and Culgoora radio spectragraph. The spectrum in the frequency range from 25 to 180 MHz is observed by Learmonth spectrograph, while the spectrum in the range of 18-25 MHz and 180-1000 MHz is by Culgoora spectrograph. The bottom panel is the same as the top panel except for the marking symbols. In the bottom panel, the data points with visible band split signals are marked by the dashed lines. Red plus and black asterisk signs delineate the F- and H-band, respectively.}
\label{fig:spectra}
\end{figure*}

In this event, the duration of the M-class flare impulsive phase is only about four minutes. The extreme fast speed and acceleration could be a combination effect of the radial motion of HC geometric center and the expansion of the HC front in all directions. The expansion of the HC front could be caused by the magnetic reconnection that adds new flux surrounding the existing component of the FR. In addition, the release of a good amount of energy in a very short time might also be the reason of the rather high value of deceleration. The deceleration is caused by the constraint of the magnetic field lines overlying the HC as will be see in Section 3.4. We find that the HC reached a distance of about 1.1 $\mathrm{R}_{\odot}$ when it attained the peak velocity at 03:24:10 UT in AIA FOV. In \citet{Zhang2006}, the analyses were based on the observations with LASCO C1, C2, and C3 whose FOV is from 1.1 $\mathrm{R}_{\odot}$. Therefore, our event can be regarded as a complementary case for the statistics of the CME kinematics. However, when comparing our event with some other events with the peak velocity also occurred below 1.1 $\mathrm{R}_{\odot}$ \citep[e.g.][]{Cheng2014}, the characteristics of its high peak velocity, acceleration and deceleration, and short energy release is still very prominent. Moreover, due to its very high velocity, although the deceleration is also very high, the HC was able to reach the LASCO C2 FOV and did not evolve to a failed-eruption event as in \citet{Cheng2014} and \citet{Song2014}.
\begin{figure*}
  \centering
  % Requires \usepackage{graphicx}
  \includegraphics[width=12. cm]{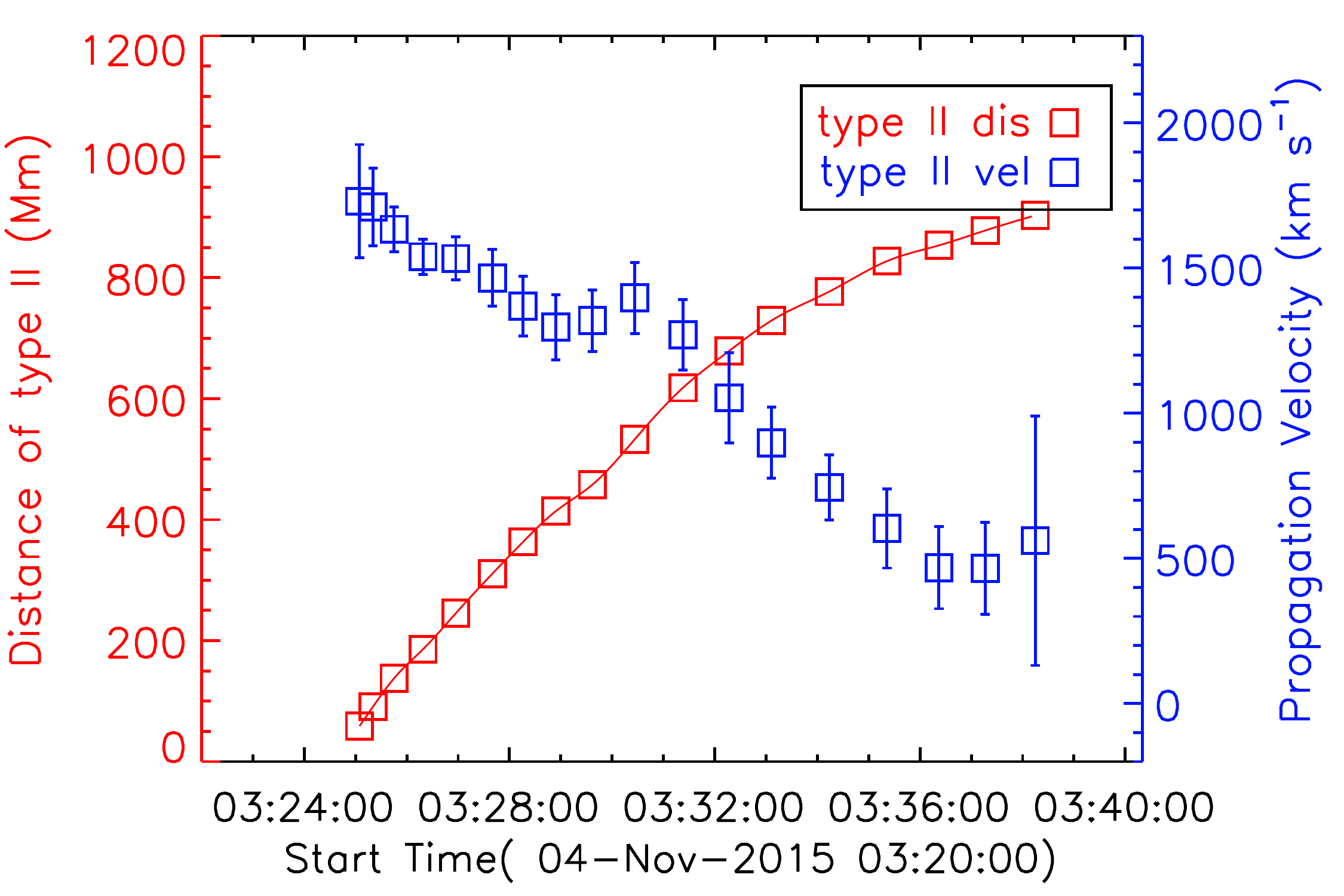}\\
  \caption{Properties of the type II bursts. The red squares represent the inverted radial distances of the type II burst using the density model of \citet{Vrsnak2004}. The blue square signs denote the shock radial speed derived from the type II radio burst.}
  \label{fig:vrs}
\end{figure*}
\begin{figure*}[htb]
\centering
\subfigure{
\begin{minipage}[t]{0.45\linewidth}
\raggedleft
 \begin{overpic}[height=0.67\textwidth,width=1.0\textwidth]{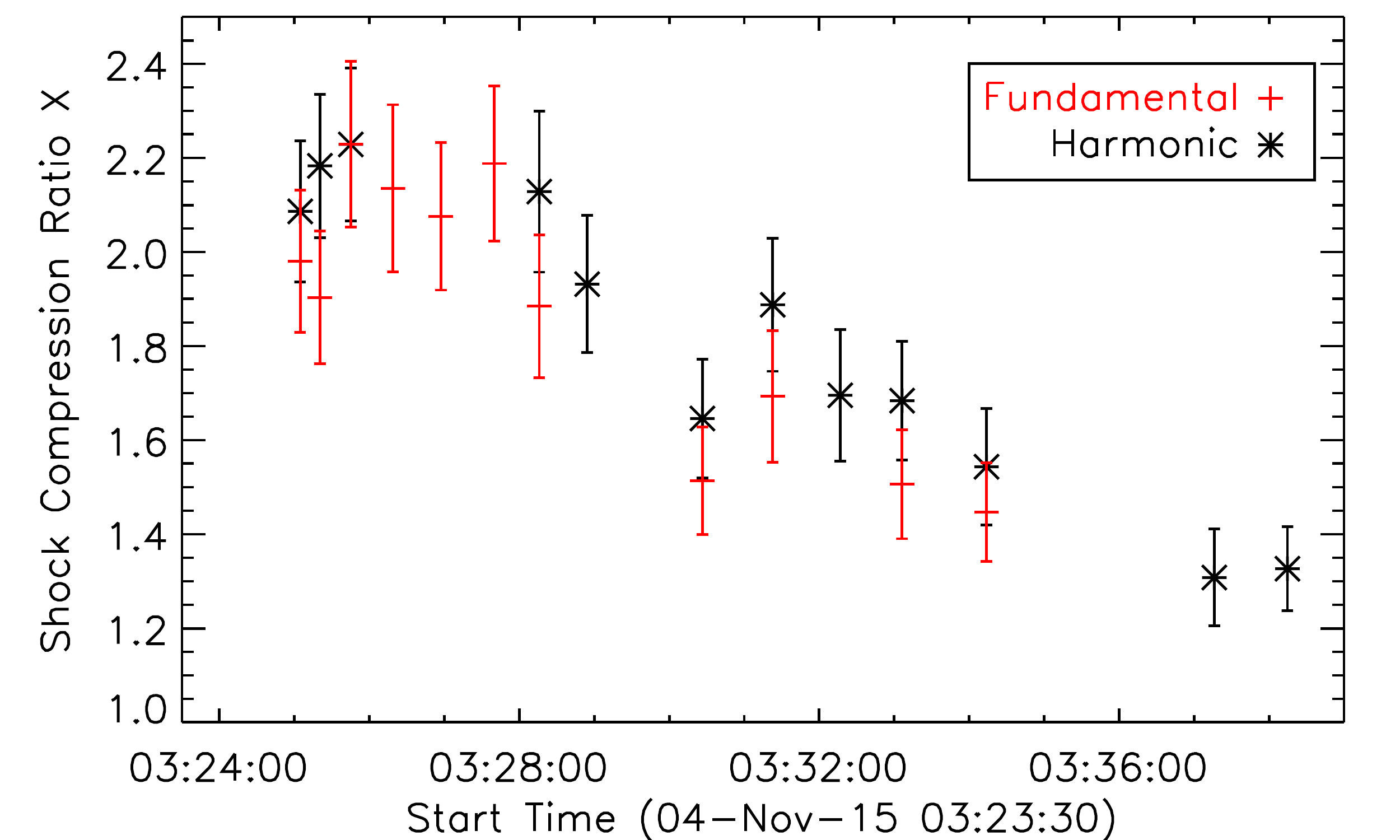}
  \put(15,12){(a)}
 \end{overpic}

 \begin{overpic}[height=0.67\textwidth,width=1.0\textwidth]{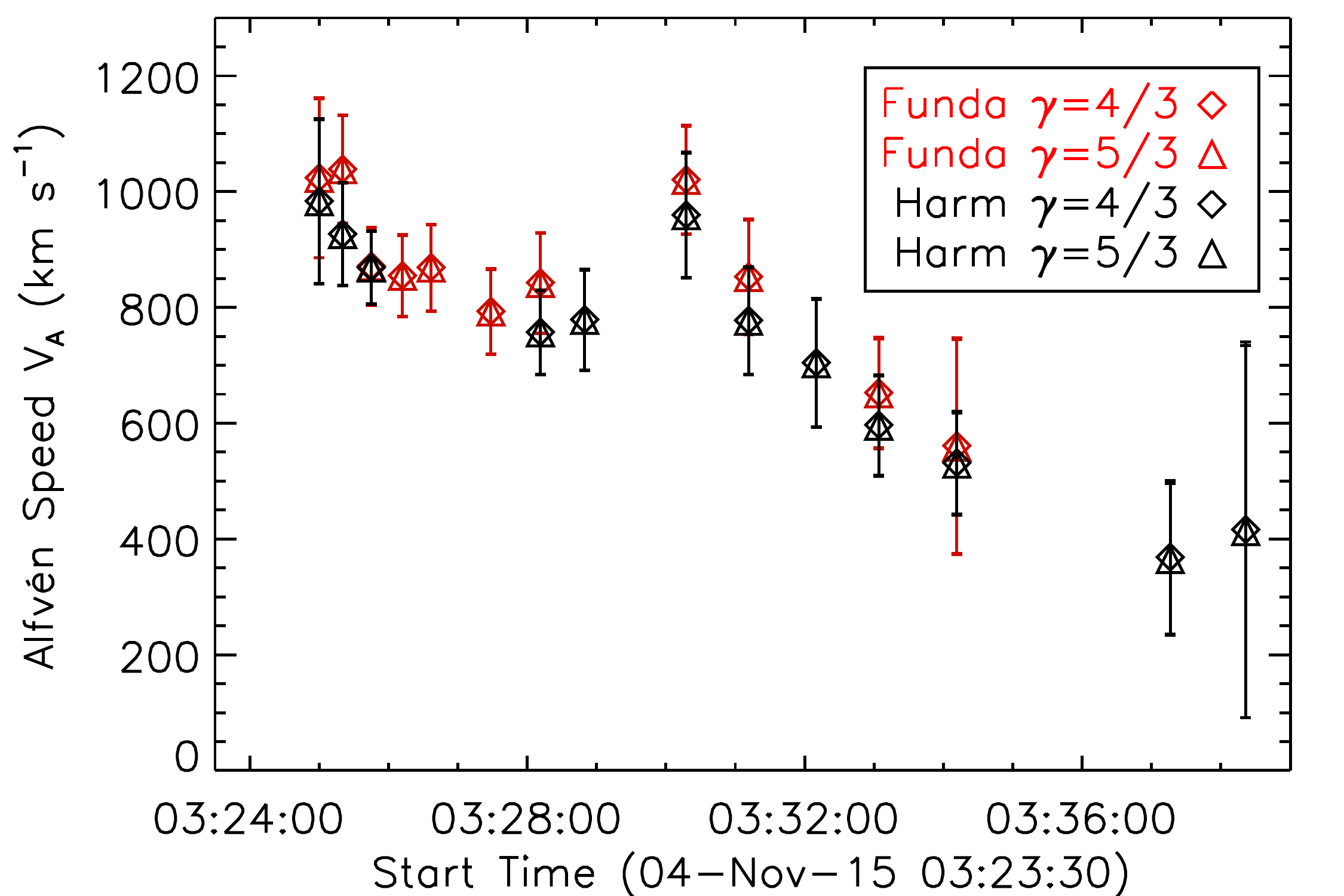}
  \put(18,13){(c)}
 \end{overpic}

 \end{minipage}
\raggedright
  \begin{minipage}[t]{0.45\linewidth}
 \begin{overpic}[width=1.0\textwidth]{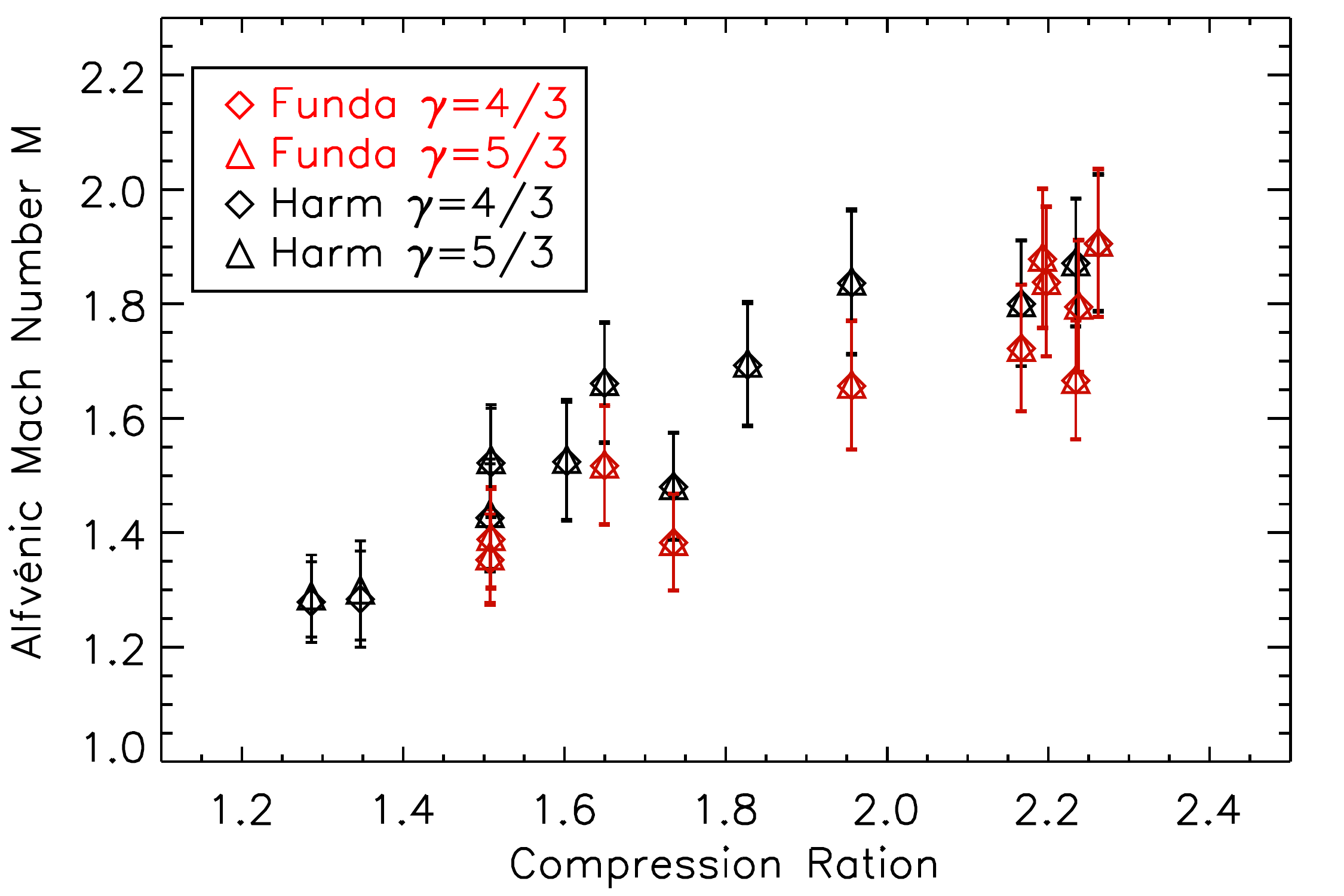}
 \put(14,13){(b)}
 \end{overpic}
 \begin{overpic}[width=1.0\textwidth]{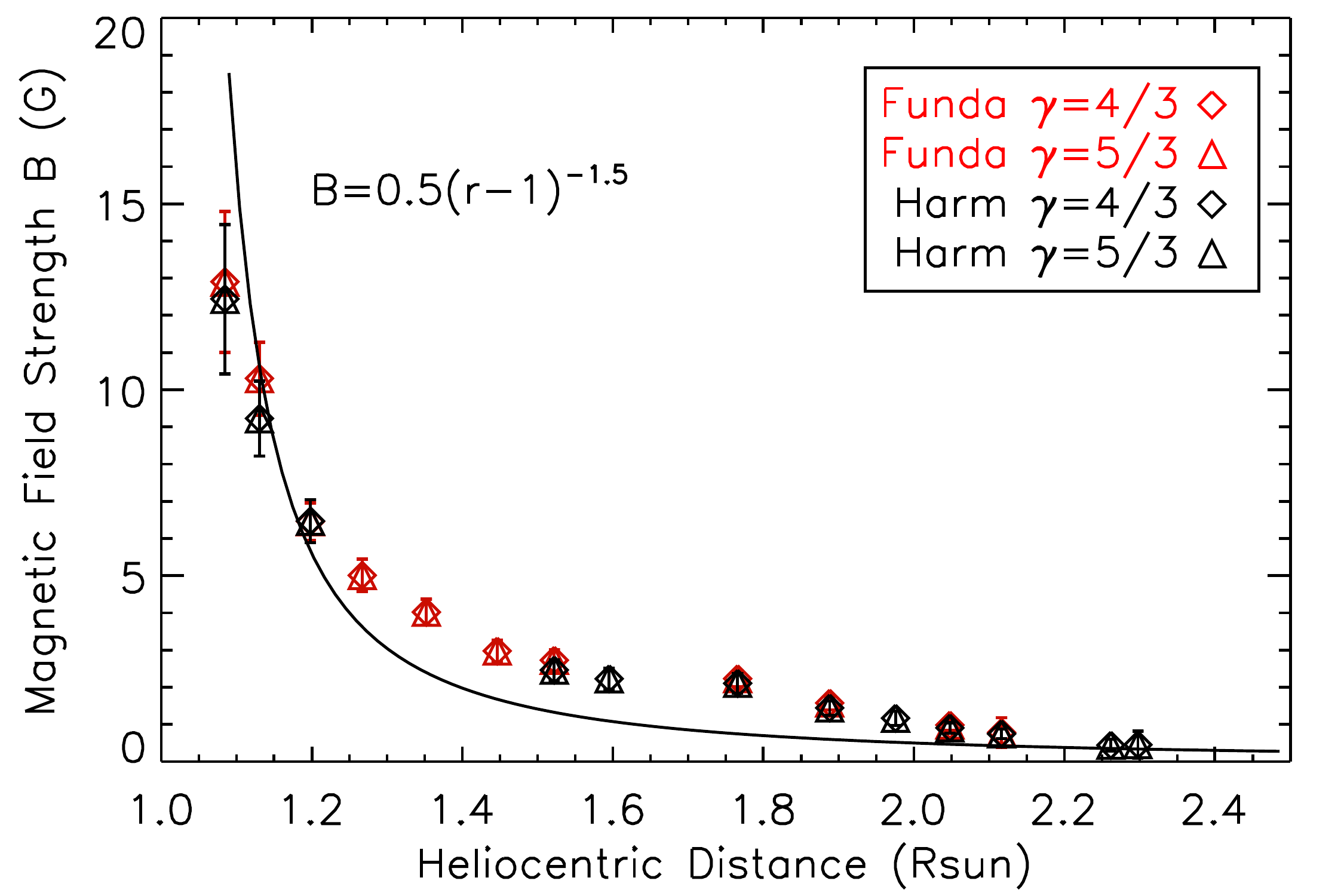}
 \put(14,13){(d)}
 \end{overpic}
   \end{minipage}
  }
\caption{(a) Shock compression ratio $X$ calculated from the band split along the F-band (red plus signs) and the H-band (black asterisks). The color code in panels (b)-(d) follow the same denotation. (b) Alfv\'{e}nic Mach number $M$ derived with the method of \citet*{Draine1993} for $\gamma = 4/3$ (diamonds) and $5/3$ (triangles). (c) Alfv\'{e}n velocity $V_{A}$ derived from the ratio of the shock radial speed $V_{sh}$ to the Alfv\'{e}nic Mach number $M$. (d) Coronal magnetic-field strength $B$ obtained from $V_{A}$ and $f_p$. The solid line is the profile derived from the empirical formula in \citet*{Dulk1978}. The error bars in all panels are derived and propagated under the assumption of 5\% uncertainty in frequency.}
\label{fig:Mach}
\end{figure*}

\begin{figure*}
  \centering
  % Requires \usepackage{graphicx}
  \includegraphics[width=1. \textwidth]{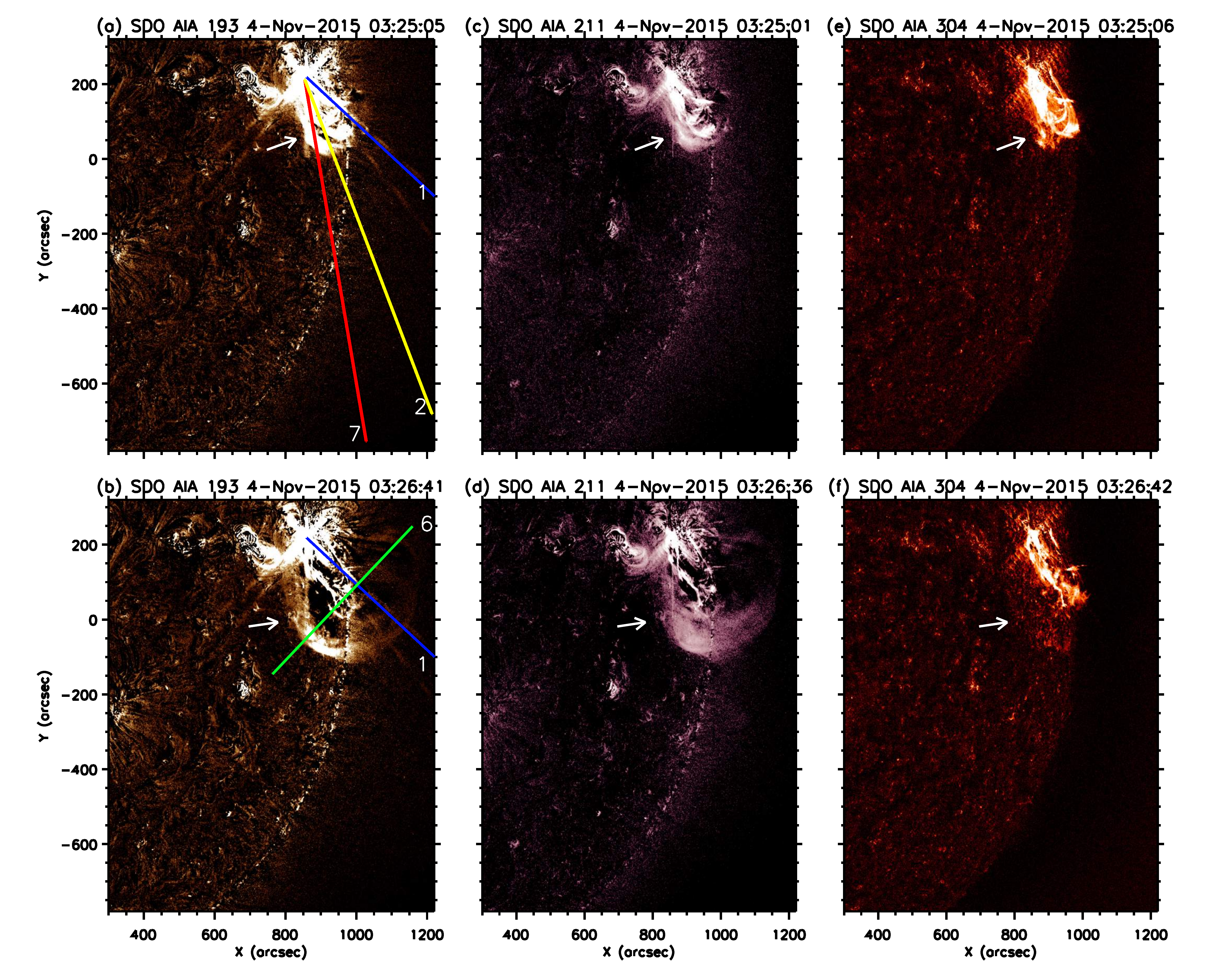}\\
  \caption{SDO/AIA base difference images in 193 {\AA} (left column), 211 {\AA} (middle column) and 304 {\AA} (right column) passbands. The time difference between the images in the first and second row is about 1.5 minutes (the time of each frame is indicated in the inset labels at the top of each panel). The slices in panel (a) numbered as 1 (in blue), 2 (in yellow), and 7 (in red) delineate the regions along which the shock kinematics is characterized  (see \autoref{fig:shock_vel}, panel a). In panel (b) the slices 1 and 6 mark the same region pointed by the homologous slices in \autoref{fig:shock} but with extended lengths. The width of the slices is 2 arcseconds.}
  \label{fig:shock_pic}
\end{figure*}

\begin{figure*}[htb]
\centering
\subfigure{
   \begin{overpic}[width=0.5\textwidth]{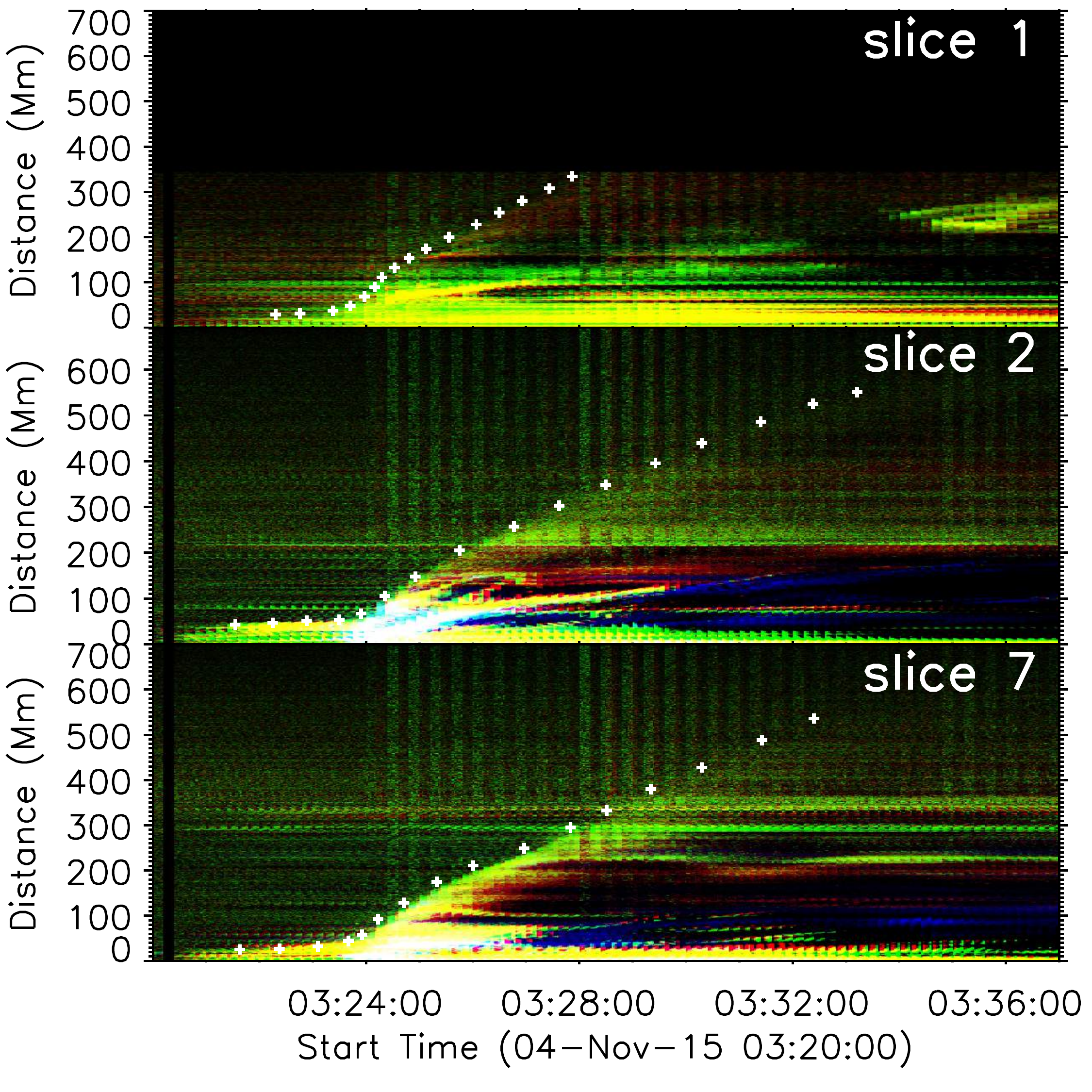}
     \put(85,15){\color{white}{(a)}}
   \end{overpic}
}
\subfigure{
   \begin{overpic}[width=0.5\textwidth]{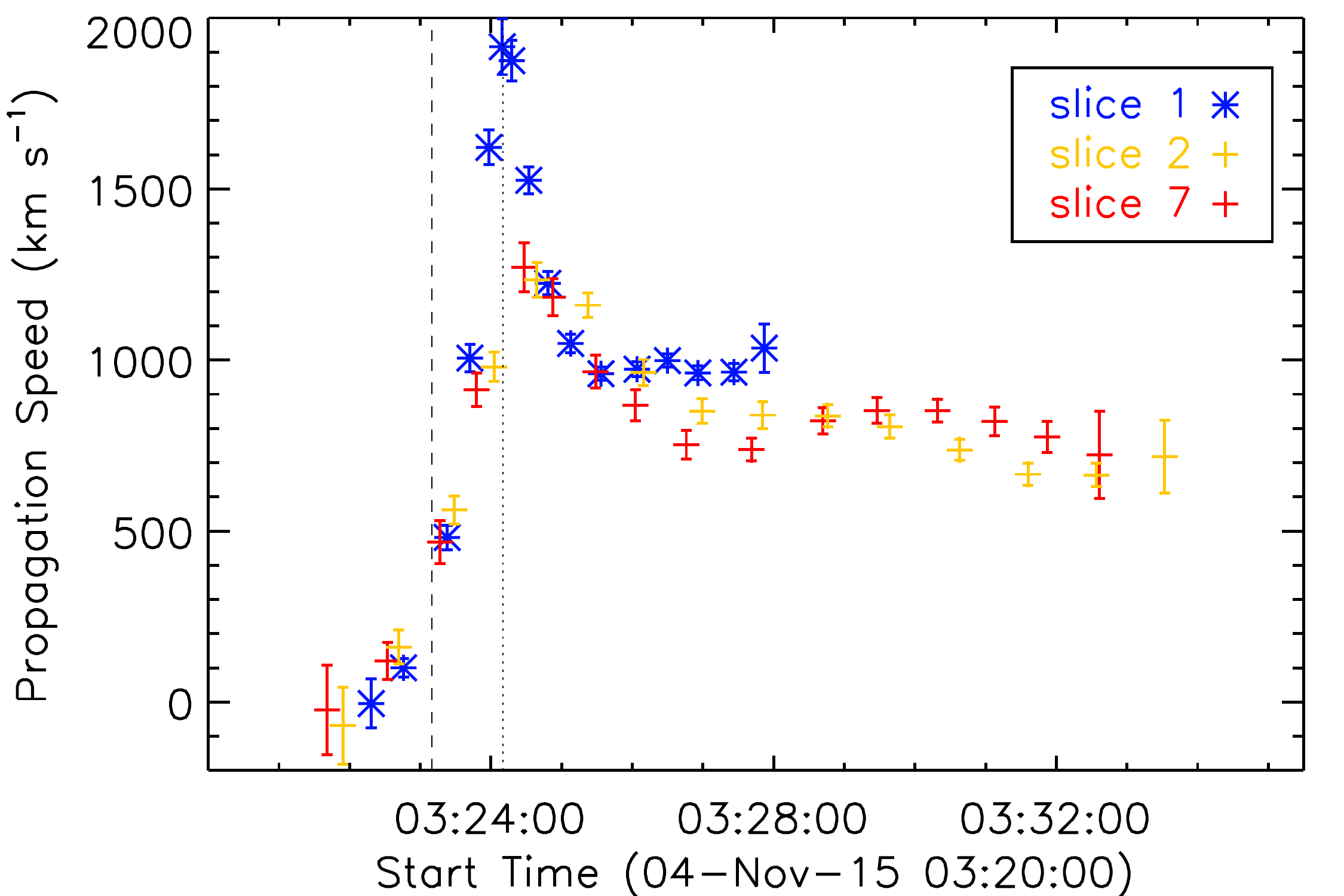}
   \put(83,13){(b)}
   \end{overpic}
}
\caption{Top panel (a): Red-Green-Blue (RGB) representation of the synthesized base difference distance-time plots along the slices numbered as 1, 2, 7 in \autoref{fig:shock_pic} (a). The emission at 193 {\AA}, 211 {\AA}, and 304 {\AA} are represented in the green, red, and blue channels, respectively. In the top panel, the composite distance-time plot only includes AIA 193 {\AA} and 211 {\AA} images, as there is almost no signal along slice 1 in 304 {\AA} images. The plus signs in white color mark the tracked positions of the bright front. Bottom panel (b): Speed of the bright front along the slices 1, 2 and 7 (same color coding as in panel a of \autoref{fig:shock_pic}; as derived from the tracks delineated by the white plus signs in the AIA tricolor frames shown in panel a). The vertical dashed and dotted line indicates the separation time between the HC and the EUV bright front along the flank and nose direction respectively shown in \autoref{fig:shock_exp}. The error bars are propagated from the uncertainty of $\pm\,6\,\mathrm{Mm}$ in the distance measurements.}
\label{fig:shock_vel}
\end{figure*}
\subsection{Piston-driven Shock Wave}

\subsubsection{Radio Type II Burst}
Type II solar radio bursts are slowly drifting structures identified in dynamic spectra. They are generally considered to be radio signatures of shock waves. Electrons accelerated at the shock front generate Langmuir waves which are converted to electromagnetic waves near the electron plasma frequency $f_p$ and $2f_p$. Type II emission bands are sometimes showing split to two lanes of similar intensity and morphology, so called band split. This characteristic is frequently consider to be due to emission from the upstream and downstream shock region \citep{Smerd1974,Smerd1975}. In our event, the metric type II burst, observed by the Learmonth and Culgoora radio spectrographs, exhibits a well defined fundamental and a second-harmonic emission band (hereinafter F- and H-band, respectively), both being split in two parallel lanes. To obtain complete F- and H-bands, we combine the dynamic spectrum observed by the Learmonth radio spectrograph in the frequency range of 25-180 MHz, and the dynamic spectrum observed by the Culgoora radio spectrograph in the frequency range of 18-25 and 180-1000 MHz. To further make the F- and H-bands prominent, we reduce the noise level. The data points in the radio-quiet time are selected as the background level of radio emission in time intervals 03:12-03:20 UT and 03:42-03:48 UT. For each frequency, we obtain a mean value of the background from the data in the chosen time ranges. Then, the intensity in each frequency is divided by the obtained mean value correspondingly. The final spectrum is presented in the upper panel of \autoref{fig:spectra}. The type II burst signals commence at about 03:25:00 UT with an unusual higher starting frequency of about 320 MHz in the F-band, which is higher than 215 MHz derived in \citet{Kumari2017} for the same event. The band-split structures are also visible which were not analyzed in \citet{Kumari2017}. The distinguished F-and H-band emissions are marked by red pluses and black asterisks, respectively. The segments with clear band-split signals are indicated by vertical dashed lines. We find that the lower (higher) H-band frequency $f_{HL}$($f_{HU}$) is proportional to the lower (higher) F-band frequency $f_{FL}$ ($f_{FU}$) with $f_{HL} \simeq 1.93f_{FL}$ ($f_{HU} \simeq 1.93f_{FU}$).

On account of the longer lifetime of the lower branch of the H-band, we utilize $f_{HL}$ to derive the upstream plasma frequency with $f_{p} \simeq f_{HL}/1.93 $, and further obtain the electron density $n$ in units of $\mathrm{cm}^{-3}$ with $f_{p} =9 \times 10^{-3} n^{\frac{1}{2}}$ MHz. To convert the derived electron density to a radial distance, a density model $n(r)$ has to be invoked. In our case, the hybrid model proposed by \citet{Vrsnak2004} is used, and it has a smooth transition from the active region corona to the interplanetary range:
\begin{equation}
\centering
n_{[10^{8} cm^{-3}]}=\frac{15.45}{R^{16}} + \frac{3.16}{R^6} + \frac{1}{R^4} + \frac{0.0033}{R^2}.
\end{equation}
The inverted radial distances are illustrated by the red squares in \autoref{fig:vrs}. We can see that the Type II source formed at a low height below 1.1 R$_{\odot}$. The red solid line is the corresponding cubic spline fitting. we have derived the velocity-time plot which is shown by blue squares in \autoref{fig:vrs}. The associated error bars are propagated from the uncertainty in frequency of $5\%$. The motion of the type II source slows down from 1750 km $\mathrm{s}^{-1}$ at 03:25 UT to 500 km $\mathrm{s}^{-1}$ at 03:38 UT.

\begin{figure*}[htb]
\subfigure{
 %\centering
  \begin{minipage}[b]{0.49\linewidth}
  \raggedleft
     \begin{overpic}[width=1.\textwidth]{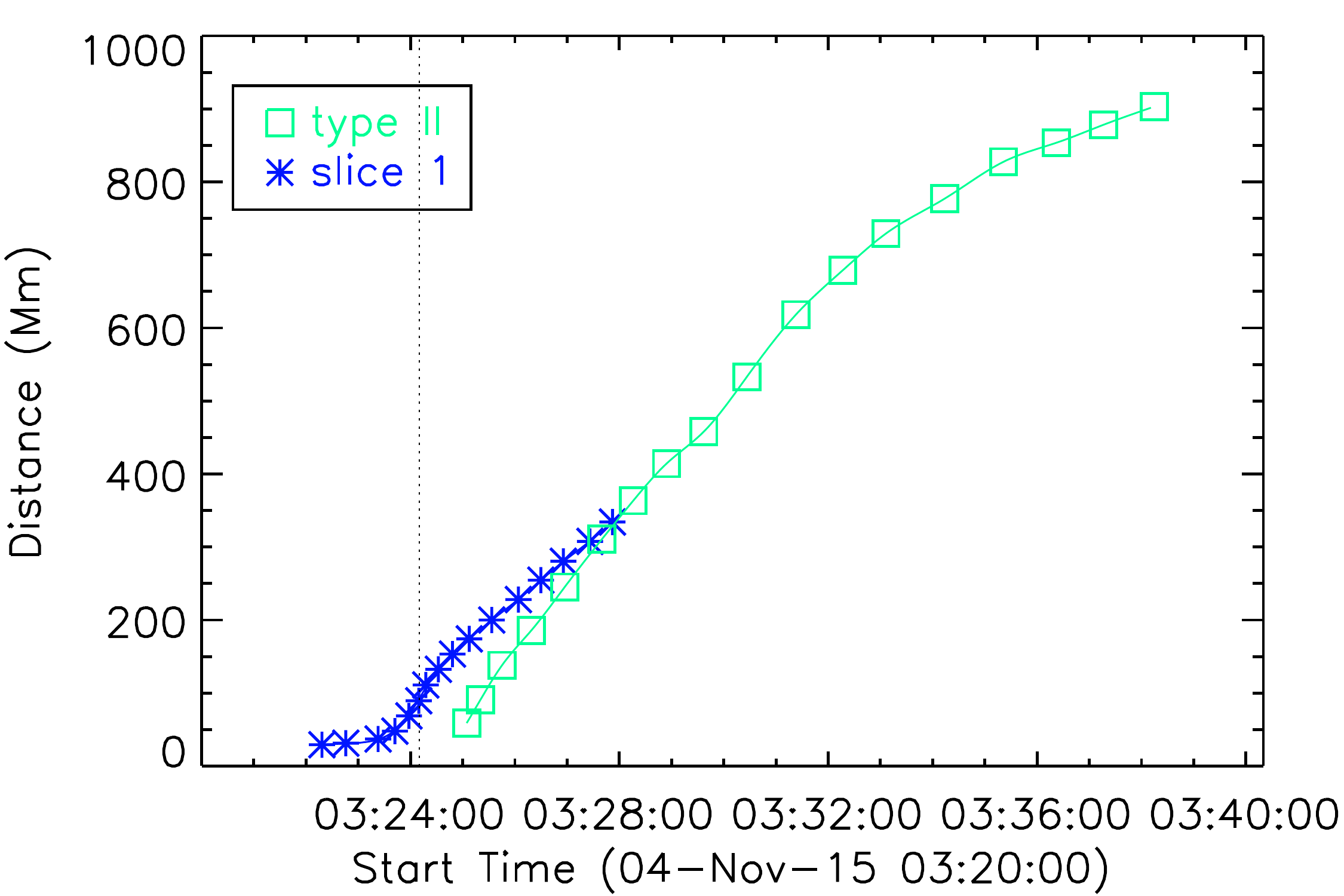}
   \put(81,12){(a)}
   \end{overpic}
  \end{minipage}

  \begin{minipage}[b]{0.49\linewidth}
  \raggedright
     \begin{overpic}[width=1. \textwidth]{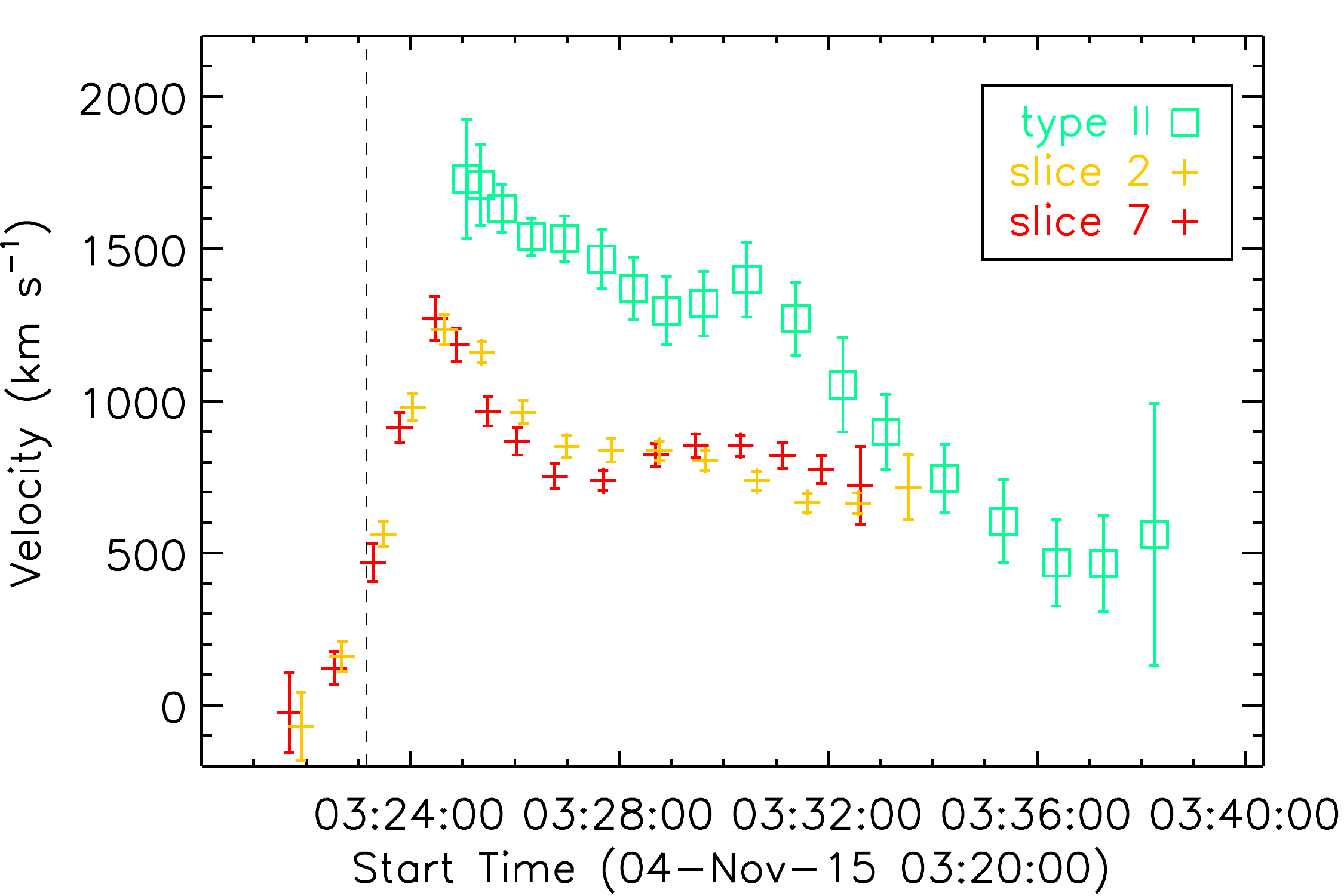}
   \put(82,12){(b)}
   \end{overpic}

  \end{minipage}
 }
 \caption{(a) Blue asterisks represent the bright front distance traced in the AIA tri-wavelength stack plot along slice 1. Green squares represent the inverted radial distances from type II burst using the density model of \citet{Vrsnak2004}. (b) Velocity evolution of the inferred shock. Green square signs denote the shock radial speed derived from the type II burst and plus signs represent the projected speeds of the shock in the AIA tri-wavelength image along slices 2 and 7. The vertical dotted (dashed) lines have the same meaning as shown in \autoref{fig:shock_vel}. The error bars of the shock radial speed (green) are derived and propagated under the assumption of 5\% uncertainty in frequency.}
\label{fig:type_vel}

\end{figure*}

Band split in radio dynamic spectra is useful to infer shock wave properties \citep{Vrsnak2001,Vrsnak2002,Vrsnak2004,Gopalswamy2012,Su2016}. The instantaneous split $\Delta f(t)$ at the moment $t$ is usually defined as the difference $\Delta f(t) = f_U(t) - f_L(t)$, where $f_U$($f_L$) is the frequency along the upper (lower) frequency branch. Then the relative instantaneous split is the ratio of $\Delta f(t)/f_L(t)$ which spans in a range from 0.14 to 0.48. The derived relative split in our event is within the range of 0.05-0.6 obtained by \citet{Vrsnak2001}. The compression ratio $X$ is defined as the shock down-stream to up-stream density ratio $n_2/n_1$, and is a quantity which characterizes the shock strength. In \autoref{fig:Mach} (a), the compression ratio, deduced with $n_{2}/n_{1} = (f_{FU}/f_{FL})^2 $ and $n_{2}/n_{1} = (f_{HU}/f_{HL})^2 $ showing red and black signs represent $X$ derived from the band split of F- and H-band, respectively. We find that the shock strength decreases from 2.2 to 1.3 within 14 minutes. To compute the Alfv\'{e}nic mach number $M$, we assume a quasi-perpendicular shock and adopt the formula introduced in \citet*{Draine1993},
\begin{equation}
%\centering
\frac{n_{2}}{n_{1}}=\frac{2(\gamma+1)}{\{D+[D^{2}+4(\gamma+1)(2-\gamma)M^{-2}]^{1/2}\}},
%n_{2}/n_{1}=\frac{2(\gamma+1)}{\{D+[D^{2}+4(\gamma+1)(2-\gamma)M^{-2}]^{1/2}\}},
\end{equation}
where $D=(\gamma-1)+(2/M_{s}^{2}+\gamma/M^{2})$, $\gamma$ is the adiabatic index and set as $4/3$ and $5/3$, and $M_{s}=V_{sh}/C_{s}$ is the sonic Mach number in which $V_{sh}$ is the radial speed of the shock presented in \autoref{fig:vrs}, $C_s$ is the sound speed. For T=1.5 MK and $m=1.27\times1.673\times10^{-24}$ g the average particle mass in the corona \citep{Aschwanden2005}, and according to the formula
\begin{equation}
\centering
C_{s} = \sqrt{\frac{\gamma p}{\rho}} = \sqrt{\frac{\gamma R  T}{M}} = \sqrt{\frac{\gamma  k  T}{m}} ,
\end{equation}
the sound speed is $C_{s} = 114$ (127) km $\mathrm{s}^{-1}$ for $\gamma = 4/3$ ($5/3$). The Alfv\'{e}nic Mach number $M$ which can be derived from Equation (2) is shown in \autoref{fig:Mach} (b). The primary uncertainty in $M$ is propagated from the compression ratio $X$. The calculated $M$ values are indicated by diamonds (triangles) with $\gamma = 4/3(5/3)$. In \autoref{fig:Mach}(c), $V_{A} = V_{sh}/M$ yields the Alfv\'{e}n speed decreasing with the heliocentric distance. Once we get $V_A$, the magnetic field strength in the upstream medium can be calculated:
\begin{equation}
\centering
B = 5.1\times10^{-5}V_{A}f_{p} .
\end{equation}
 The upstream frequency $f_{p}$ has been obtained thereinbefore, and shown in \autoref{fig:Mach} (d). \citet*{Dulk1978} derived an analytical equation for $B(r)$:
\begin{equation}
\centering
B = 0.5(r-1)^{-1.5},
\end{equation}
where $r$ is the heliocentric distance in units of solar radii. For $r(R_{\odot})=1.3-2.1$, our measurements are somewhat higher, but comparable with the value given by this formula.

\begin{figure*}[htb]
%\centering
%\raggedright
\begin{minipage}[b]{0.49\linewidth}
  \raggedleft
\subfigure{
  %\begin{minipage}[b]{6.0 cm}
  %\raggedright
  %\hspace{-0.1 cm}
   %\begin{overpic}[width=0.49\textwidth,height=0.31\textwidth,clip]{hotchanel_exp_comb_slice1.eps}
   \begin{overpic}[width=1.\textwidth,clip]{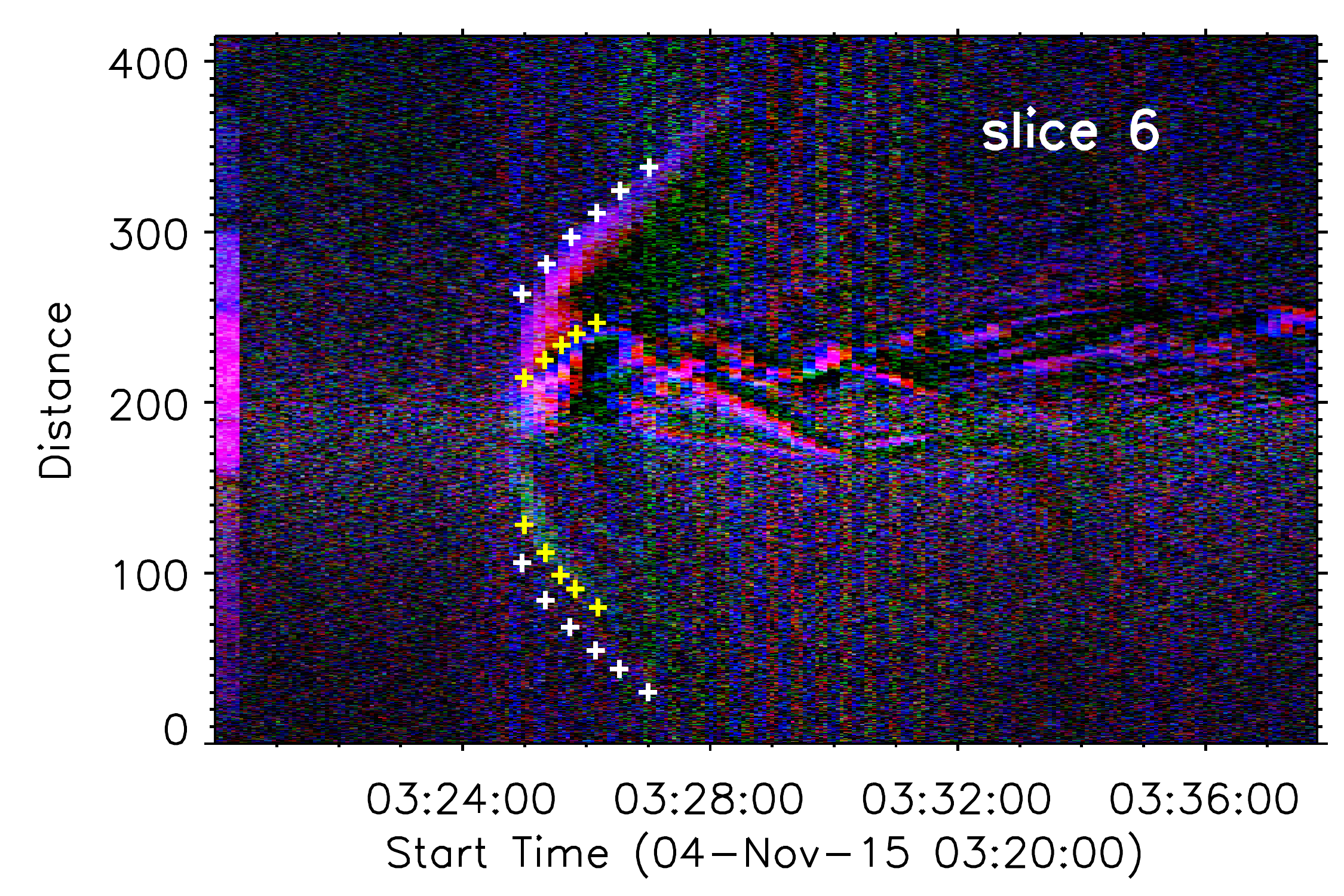}
     \put(86,14){\color{white}{(a)}}
   \end{overpic}
  %}

 % \subfigure{
   %\begin{overpic}[width=0.49\textwidth,height=0.31\textwidth,clip]{hotchanel_exp_comb_slice1.eps}
      \begin{overpic}[width=1.\textwidth,]{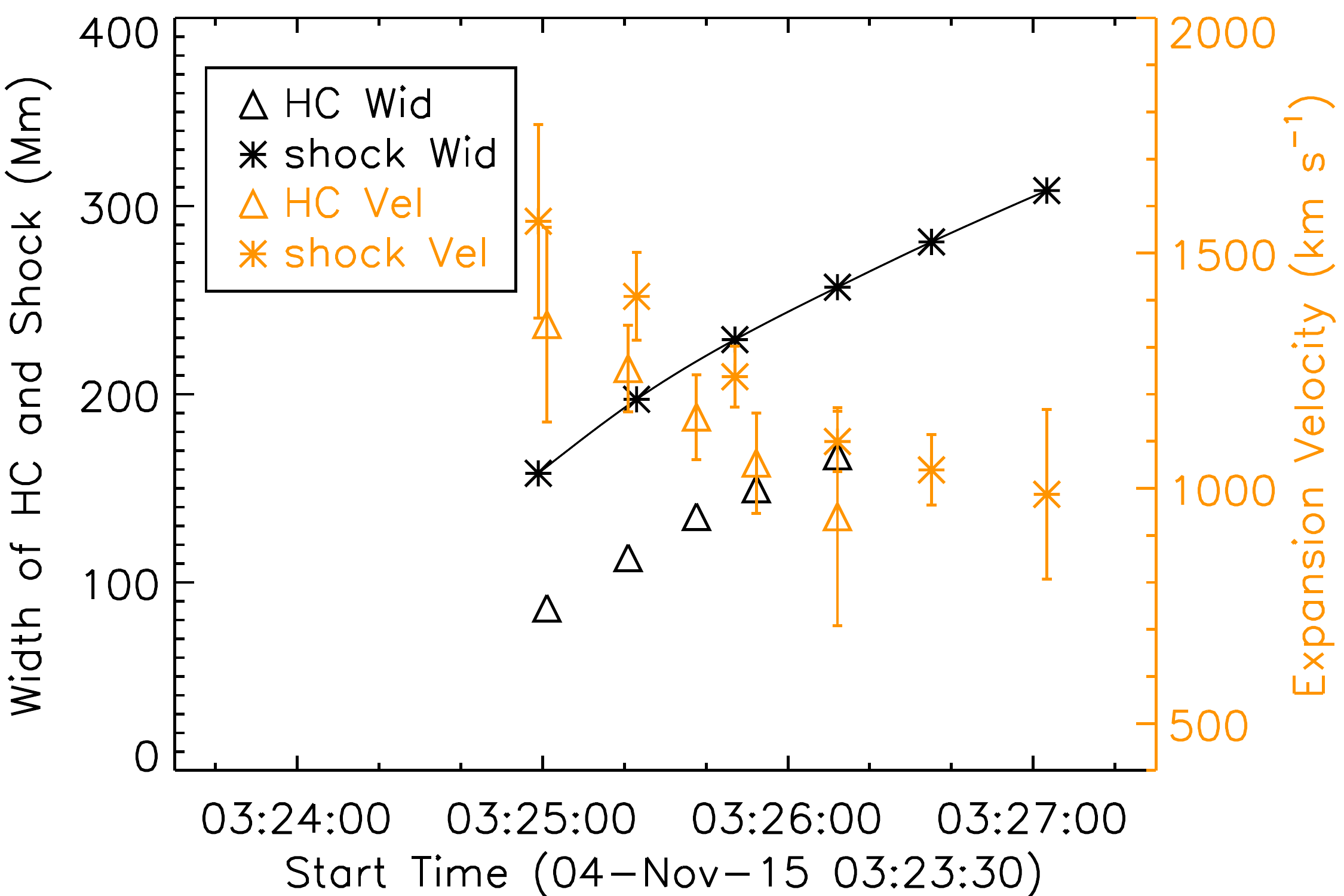}
   \put(76,14){(d)}
   \end{overpic}

  }

\end{minipage}

\begin{minipage}[b]{0.49\linewidth}
  \raggedright
  \subfigure{
   %\begin{minipage}[b]{6.0 cm}
   %\raggedright
         %\begin{overpic}[width=0.49\textwidth,,height=0.29\textwidth]{exp_vel_dis_comb.eps}

    \begin{overpic}[width=1.\textwidth,clip]{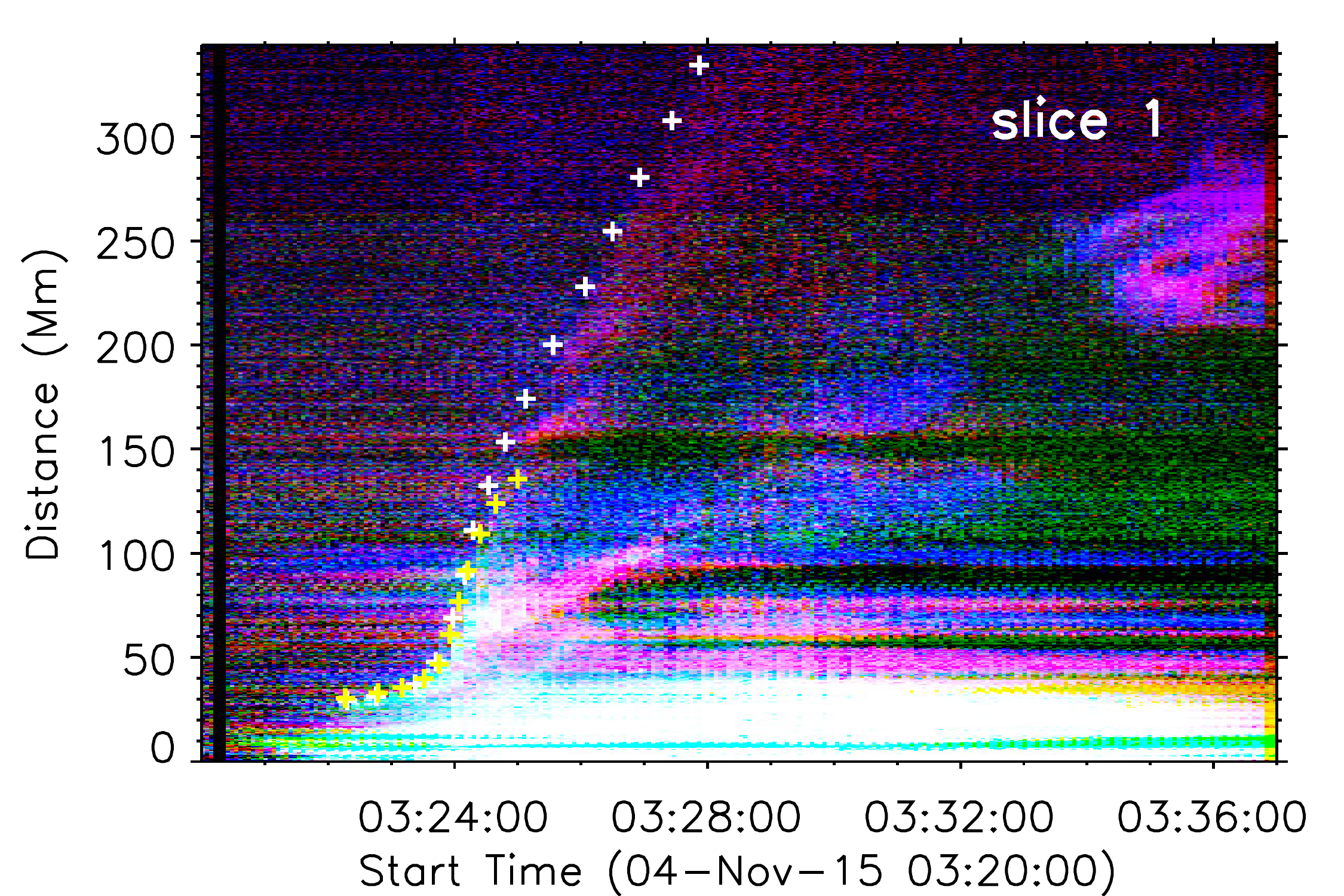}
     \put(86,14){\color{white}{(b)}}
   \end{overpic}

%}

%\subfigure{

   % \begin{overpic}[width=0.49\textwidth,height=0.3\textwidth]{prop_94_193_1.eps}
   \begin{overpic}[width=1.\textwidth]{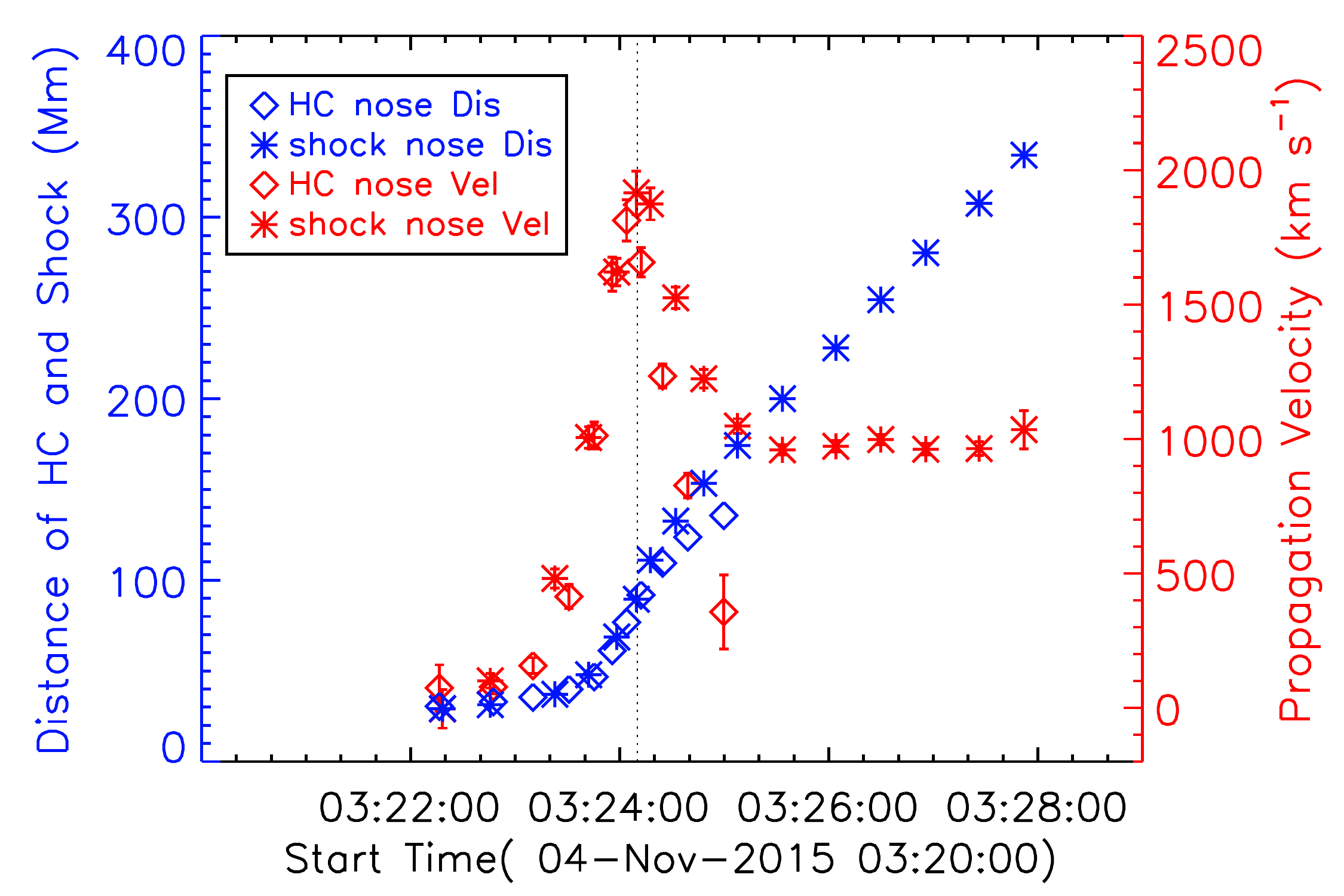}
   \put(76,14){(e)}
   \end{overpic}

}
\end{minipage}

\begin{minipage}[b]{0.49\linewidth}
  \raggedright
  \subfigure{
   %\begin{minipage}[b]{6.0 cm}
   %\raggedright
         %\begin{overpic}[width=0.49\textwidth,,height=0.29\textwidth]{exp_vel_dis_comb.eps}

    \begin{overpic}[width=1.\textwidth,clip]{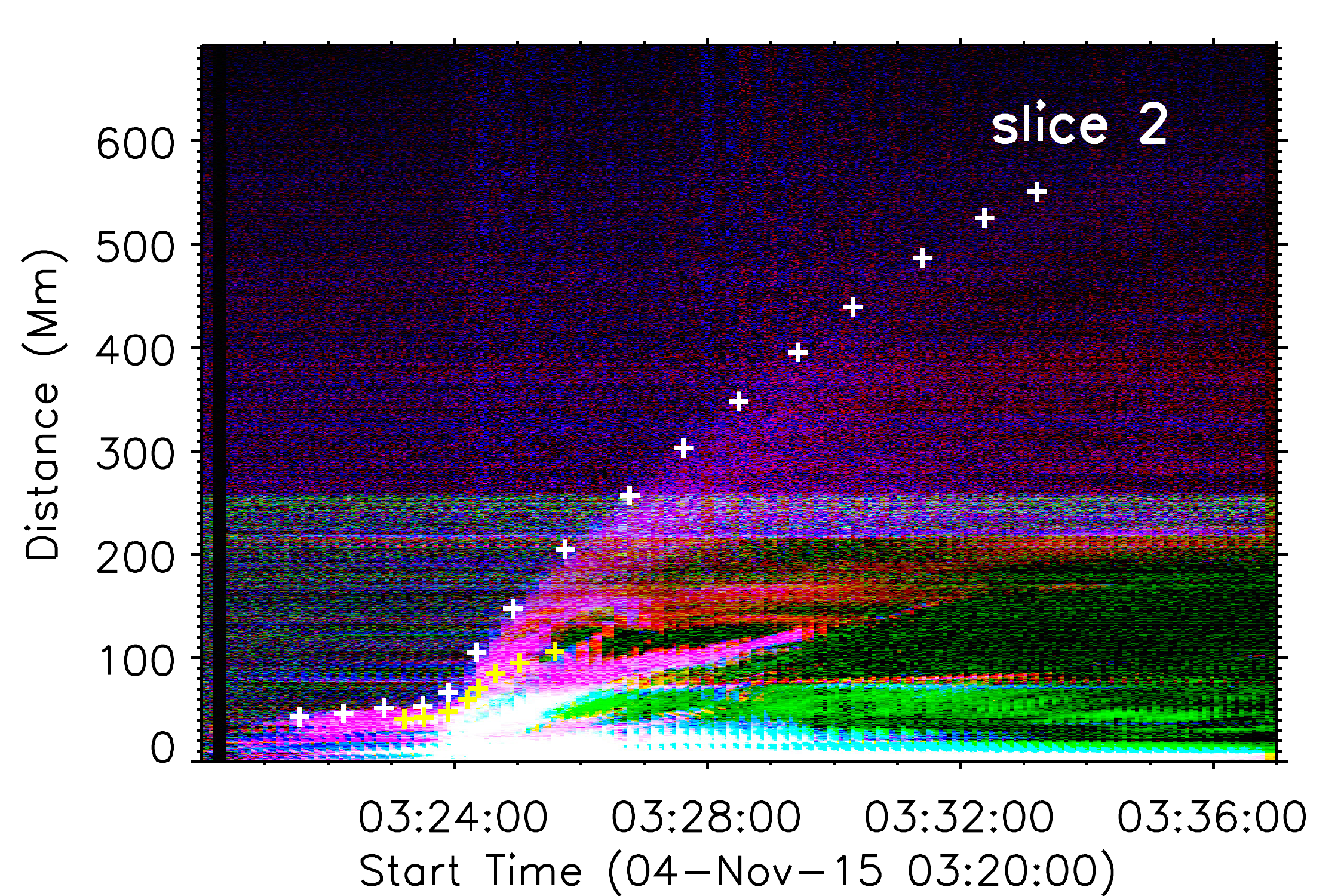}
     \put(86,14){\color{white}{(c)}}
   \end{overpic}

%}

%\subfigure{

   % \begin{overpic}[width=0.49\textwidth,height=0.3\textwidth]{prop_94_193_1.eps}
   \begin{overpic}[width=1.\textwidth]{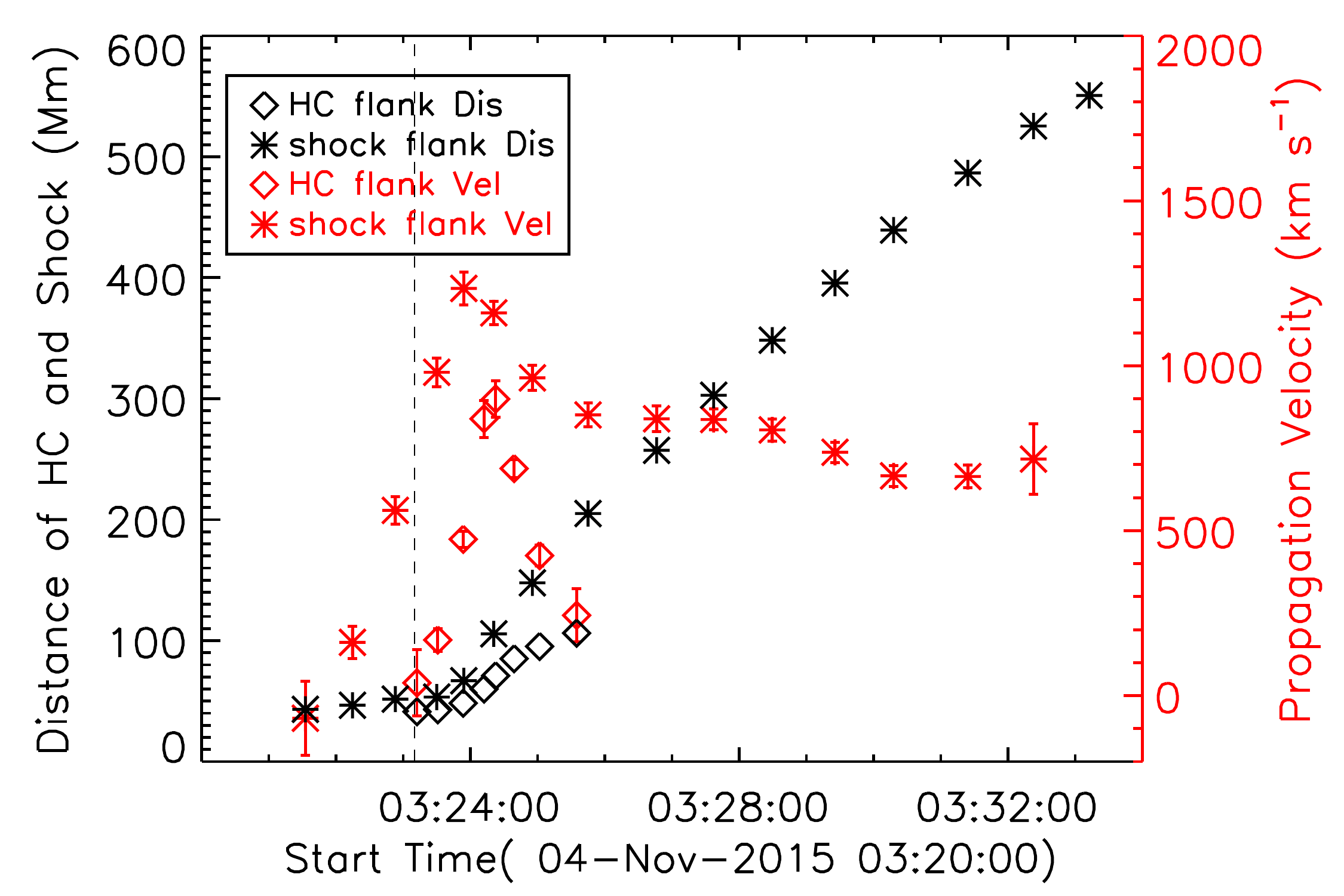}
   \put(76,14){(f)}
   \end{overpic}

}
\end{minipage}
\caption{Panel (a): RGB representation of the synthesized running difference distance-time plot along slice 6 with the 211~{\AA} image in the red channel, the 193~{\AA} image in the blue channel and the 94~{\AA} image in the green channel. Panels (b)-(c): RGB representation of the synthesized base difference distance-time plots along slices 1 and 2. In panels (a-c), the shock and HC fronts are denoted by white and yellow plus signs, respectively. Panel (d): Lateral expansion width of the HC (black triangles) and of its associated shock (black asterisks) along slice 6 (scale on the left axis), along with their corresponding lateral expansion velocity (in orange color, scale on the right axis) as derived from the smoothing cubic-spline fittings to the width-time data points. Panel (e): Distance-time (in blue color, scale on the left axis) and velocity-time (in red color, scale on the right axis) plots of the HC (diamonds) and the shock (asterisks) along slice 1. Panel (f): Evolution of the distance (scale on the left axis) and velocity (scale on the right axis) of both the HC and the shock along slice 2. The dotted (dashed) lines indicate the time when the shock and the HC along slice 1 (slice 2) start to separate in space.}
\label{fig:shock_exp}
\end{figure*}
\subsubsection{Imaging Observations}

\begin{figure*}[htb]
%\raggedright
\centering
%\begin{overpic}[scale=.5,grid,tics=10]%
 %             {20151104_shock_hotchannel.eps}
%\end{overpic}
%\hspace{-1. cm}
%\subfigure{

  \begin{overpic}[width=0.7 \textwidth]{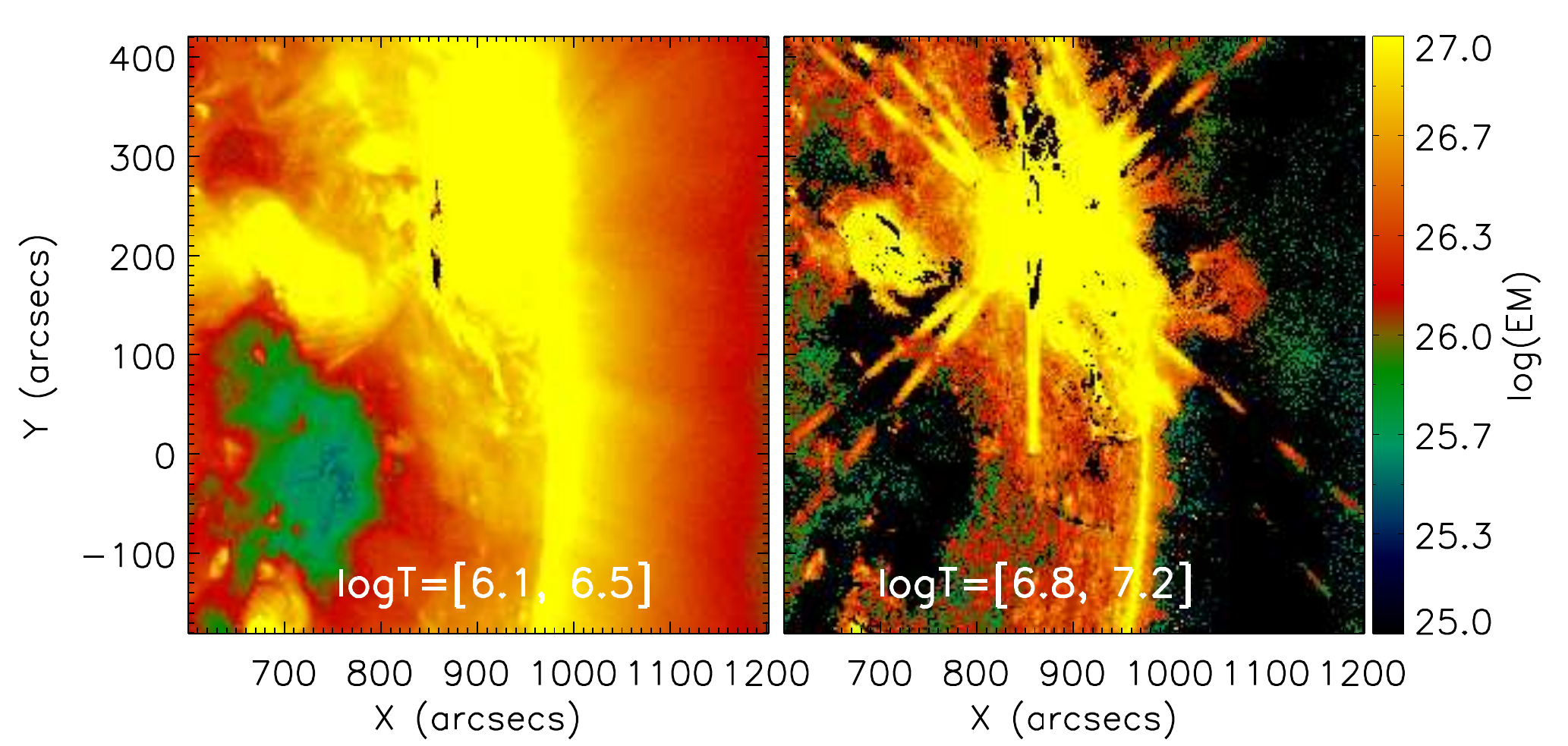}
  \put(42,42){\color{white}{(a)}}
  \put(81,42){\color{white}{(b)}}
  \put(22,20){\large\color{blue}$\Rightarrow$}
  \put(70,17){\large\color{blue}$\Uparrow$}
   \end{overpic}

   \centering
          \begin{overpic}[width=0.42 \textwidth]%
               {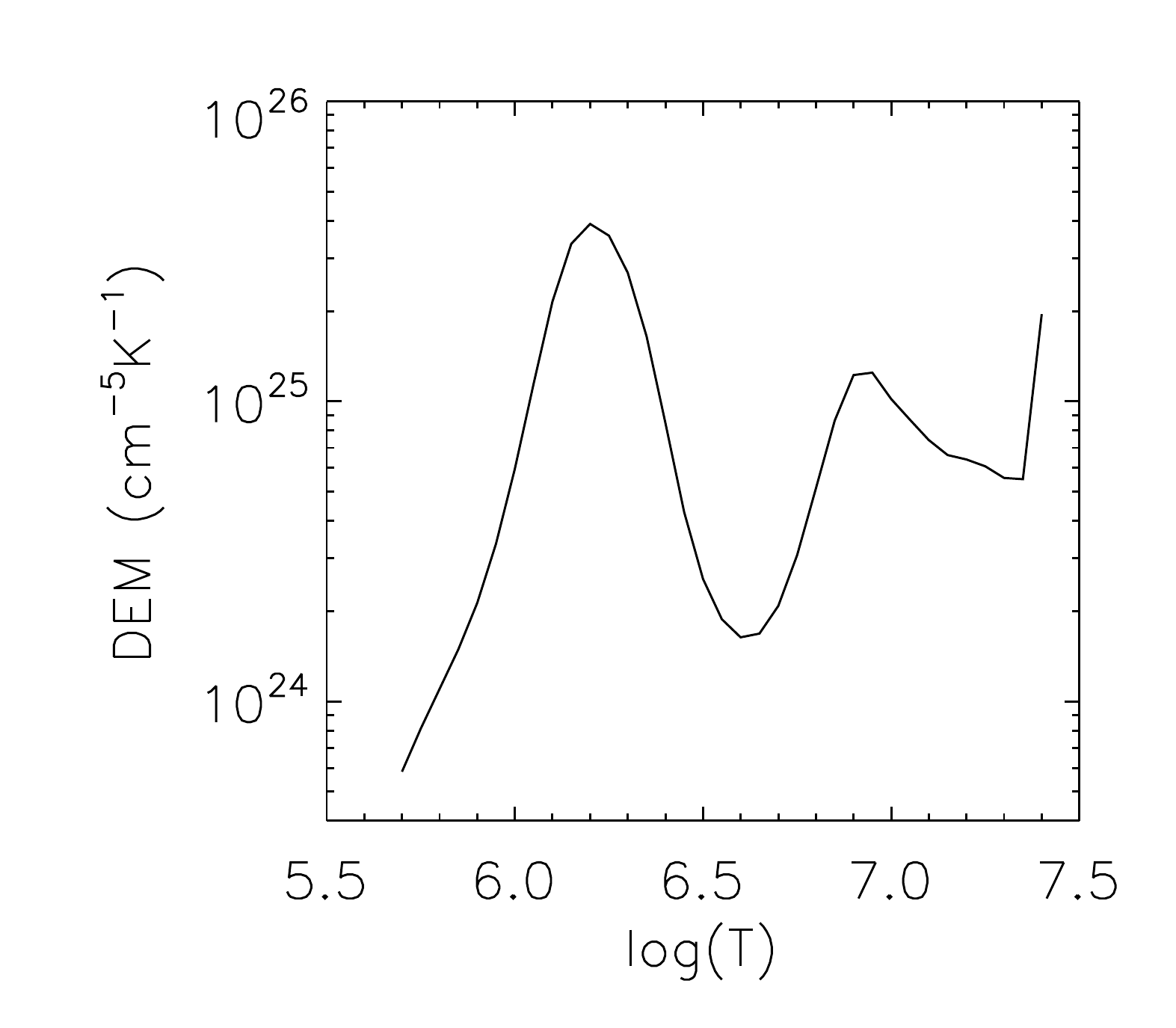}
        \put(50,70){1}
        \put(73,57){2}
        \put(83,73){(c)}
    \end{overpic}

   % }
%\includegraphics[width=14.cm]{20151104_shock_hotchannel.eps}
\caption{(a) EM in temperature range from $6.1-6.5$ in logarithm scale in which the presumed shock structure is best seen. The blue arrow points to the position of the shock region. (b)  EM in the temperature range from $6.8-7.2$ in logarithm scale in which the hot channel is best seen. The blue arrow denotes the position of the HC. (c) The sum of the DEM curve of each pixel in the region shown in panels (a) and (b). The first peak corresponds to the contribution of the emission mainly from the shock compression region, while the second peak from the HC.
}
\label{fig:Temper}
\end{figure*}
As indicated in \autoref{fig:HC}, a bright front appeared preceding the HC. \autoref{fig:shock_pic} shows the evolution of the EUV bright front in AIA multiple wavelengths. The top images of \autoref{fig:shock_pic} is observed at $\sim$03:25 UT at which the shock signal first appeared in radio dynamic spectra. The bright front is clearly seen in AIA 193 {\AA} and 211 {\AA} images, and only part of it can be distinguished in AIA 304 {\AA} image at $\sim$03:26:40 UT. \autoref{fig:shock_vel} (a) is the composite of the tri-color AIA images in 193 {\AA} (green channel), 211 {\AA} (red channel) and 304 {\AA} (blue channel) along slices in \autoref{fig:shock_pic} (a). Note that due to lack of the bright front signature in AIA 304 {\AA} along slice 1, the first panel in \autoref{fig:shock_vel} (a) is a bi-color image in AIA 193 {\AA} and 211 {\AA}. Here, we regard the direction along slice 1 as the nose direction. The distance and derived velocities along slices 1, 2 and 7 are shown in \autoref{fig:shock_vel} (a) and (b). As indicated in \autoref{fig:shock_vel} (b), the bright front experiences an acceleration phase and reaches a maximal velocity of about 1900 km $\mathrm{s}^{-1}$ at 03:24:10 UT and then decelerates. The vertical dashed line indicates the time when the bright front separates from the HC flank, while the dotted line denotes the separation between the HC nose and the bright front in space. At 03:25 UT, although the speed of the bright front decreases to around 1200 km $\mathrm{s}^{-1}$, it still reaches over the local Alfv\'{e}n speed shown in \autoref{fig:Mach} (c). In \citet{Kumari2017} the Gauribidanur RAdioheliograPH (GRAPH) observations revealed that the type II source at 03:31~UT was located immediately above the EUV bright front observed at 03:26~UT. Therefore, in combination of the GRAPH source location, analyses of the type II burst in radio dynamic spectra, and EUV bright front evolution, we tend to believe that these bright fronts have close relationship with the shock fronts both in time and in space. At least after 03:25~UT we can say that the bright fronts have gone through the nonlinear evolution and transformed into a shock wave. Before 03:25~UT the bright fronts may simply be an EUV wave front.
\begin{figure*}[t]
\centering
\subfigure{
\begin{overpic}[bb=10 18 580 300, width=0.6\textwidth,clip]{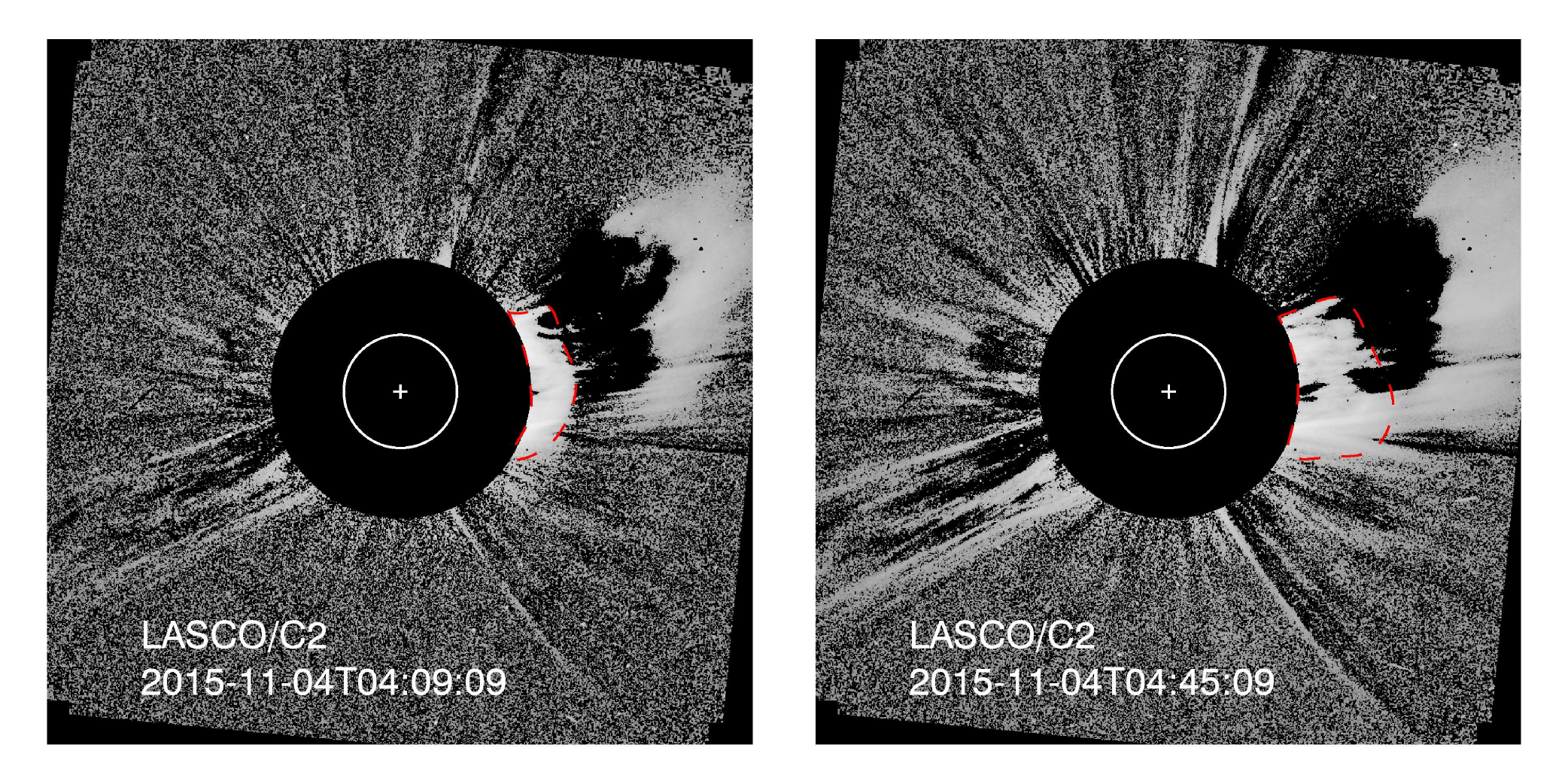}
\put(44,40){\LARGE\color{white}{$\Downarrow$}}
\put(5,43){\color{white}{(a)}}
\put(55,43){\color{white}{(b)}}
\end{overpic}

\begin{overpic}[width=0.35\textwidth,height=0.283\textwidth]{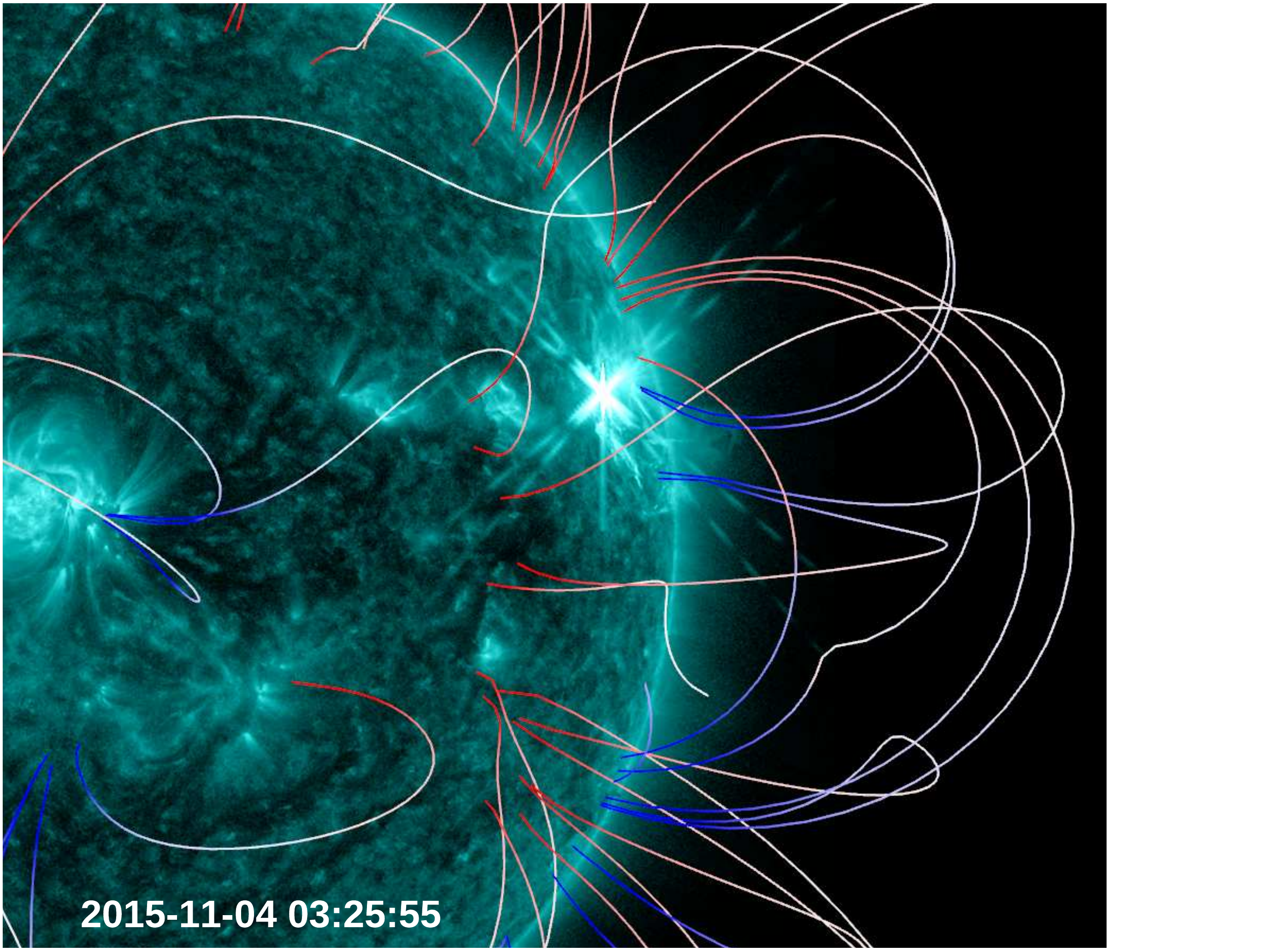}
\put(47,52){\huge\color{red}{$\star$}}
\put(47,12){\huge\color{red}{$\star$}}
\put(5,73){\color{white}{(c)}}
\end{overpic}
}
%\vspace{0.25 cm}
%\includegraphics[width=14.cm]{cme.eps}
\caption{(a)-(b) White-Light coronagraph observations of the CME at different times. The red dashed line defines the leading front of the CME, and the white arrow points to a preceding CME event. The white circle and plus sign defines solar disk and solar center, respectively. (c) Coronal magnetic field lines extrapolated with the PFSS method over-plotted in the AIA 131~{\AA} image. The footpoints of the streamer arcades overlying the HC are marked by two red stars.}

\label{fig:cme}
\end{figure*}

\begin{figure*}[htb]
\centering

 \begin{overpic}[width=0.9\textwidth]{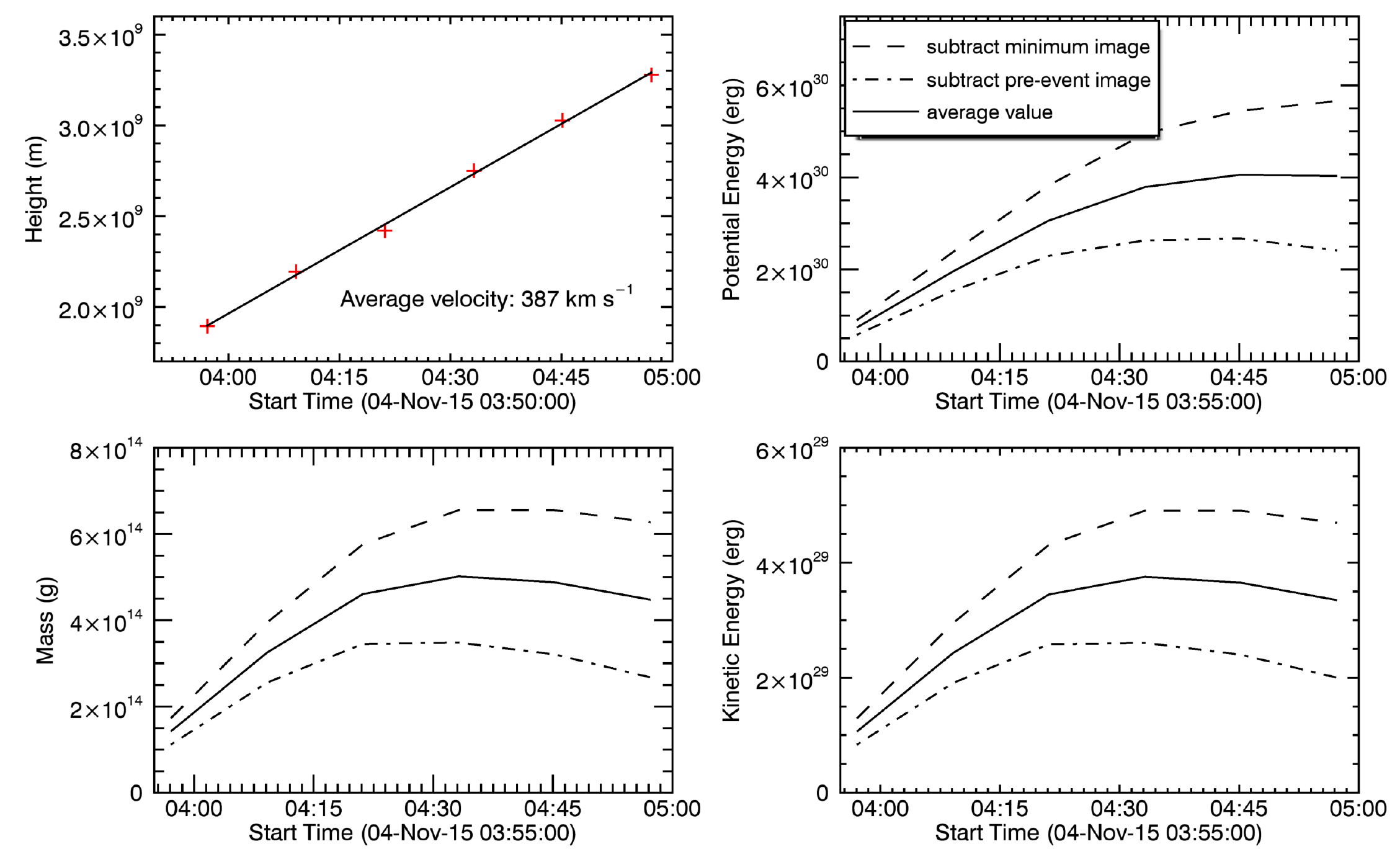}
  \put(43,37){(a)}
  \put(43,6){(b)}
  \put(92,37){(c)}
  \put(92,6){(d)}
 \end{overpic}

\caption{Panel (a): Height-time plot of the CME leading edge. Panels (b)-(d): Mass, potential energy, and kinetic energy  profiles, respectively, of the CME feature. The dashed lines represent the upper limit, which was obtained from subtracting a 12-hour minimum image; and the dotted lines depict the lower limit, which was obtained from subtracting a pre-event image as the background. The average values of the upper and lower limits are denoted by the solid lines.}
\label{fig:cme_vel}
\end{figure*}

\begin{figure}[htb]
\centering
    \begin{overpic}[width=8.5 cm]%
               {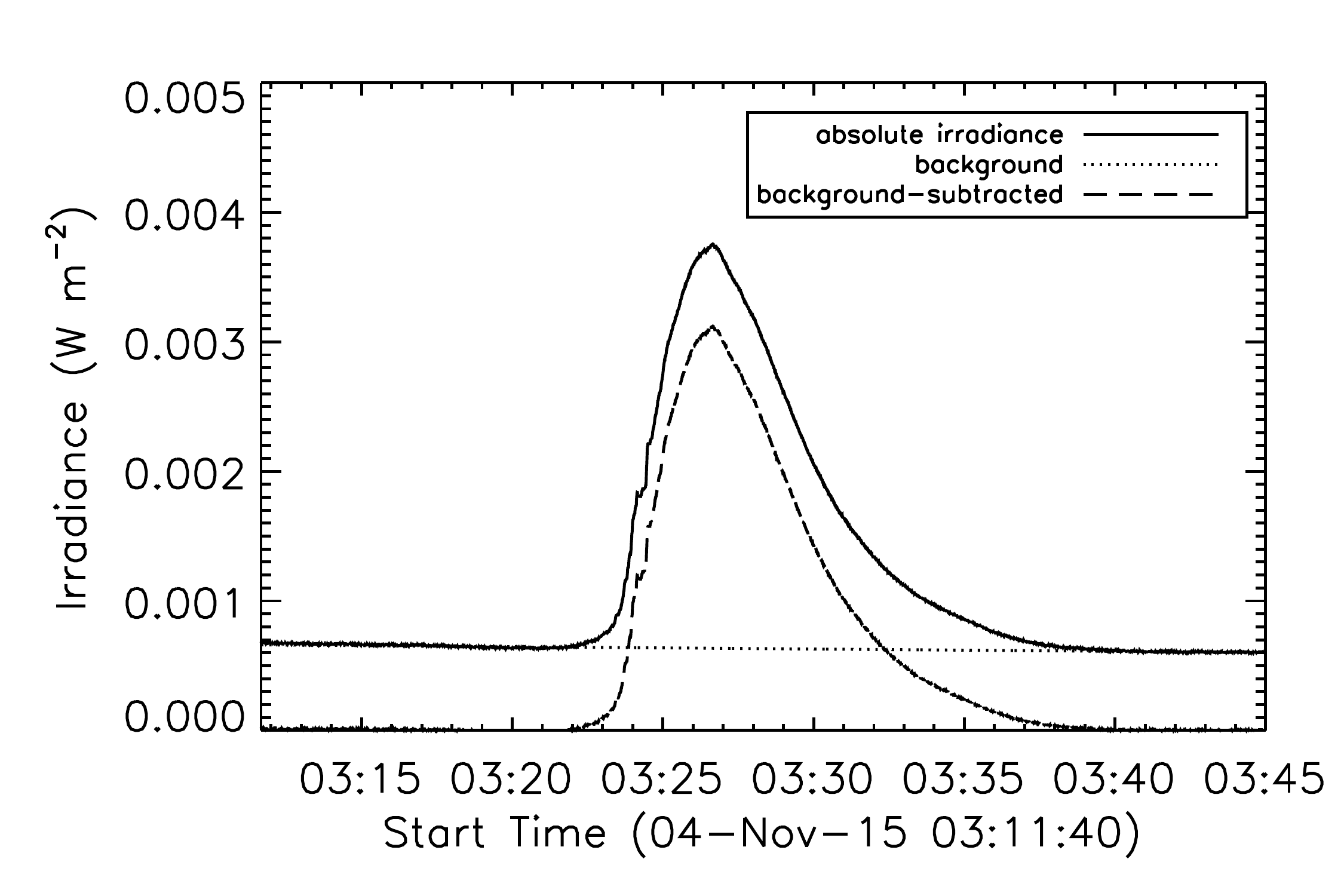}

    \end{overpic}
 \caption{The irradiance in the range of 0.1-7 nm recorded by the SDO EVE/ESP. The solid line presents the absolute irradiance. The background and background-subtracted irradiance  are delineated by dotted line and dashed line, respectively.}
    \label{fig:eve}
\end{figure}

In \autoref{fig:type_vel} (a), the inverted radial distances from the type II radio burst are illustrated by the green squares, together with the bright front distance marked by the blue asterisks, which are traced in the AIA composite image along the direction of the nose. The blue and green solid lines are the corresponding cubic spline fittings. In \autoref{fig:type_vel} (b), the distinction between the velocity derived from type II frequency drift and the velocities from AIA imaging observations can be seen. \autoref{fig:type_vel} (a) and (b) confirm non-radial propagation of the shock wave. Namely, the coronal electron density models describe the change of density in radial direction, and they are not very suitable for events with strongly non-radial propagation. For the type II source, the radial distances are calculated; while for the presumed shock in AIA images, the distances are measured along a given non-radial direction in the solar disk plane.

We have also investigated the stand-off distances between the HC and its driven shock along slices 1, 2 and 6. \autoref{fig:shock_exp} (a-c) are composite distance-time images along these three slices with red, green, and blue channels in AIA 211 {\AA}, 94 {\AA}, and 193 {\AA}, respectively. The traced shock and HC fronts are denoted by white and yellow plus signs. The measured widths of the HC and the shock along slice 6 are indicated in. \autoref{fig:shock_exp} (d) with black triangles and asterisks, respectively. It can be found that the stand-off distance increases with time in this direction. The expansion velocities of the HC are slightly lower than the velocities of the shock as demonstrated by the orange triangles and asterisks.\autoref{fig:shock_exp} (e) includes the distance-time and velocity-time plots of the HC and the shock along the nose direction (slice 1) marked by blue and red symbols, respectively. The propagation velocity of the shock is also higher than the HC's velocity. The vertical dotted line indicates the time 03:24:10 UT, when the bright front and the HC started to separate in space. \autoref{fig:shock_exp} (f) reveals similar relation between the HC and the shock front along the flank direction (slice 2), and the separation time is $\sim$1 minute earlier at~03:23:10 UT. We assume that the shock formation time is no later than the time when the shock signature was captured by radio spectra at around 03:25 UT. Thus, the formation time of the shock is less than 2 minutes after the onset of the HC main acceleration phase. Before 03:25 UT, the shock nose reaches a distance less than about 180 Mm. The stand-off distance between the HC nose and the shock nose slightly increases with time.

The similarity in the overall evolution of the HC and the shock and the increase of the stand-off distance with time fit to the piston-driven shock scenario \citep{Warmuth2015}. A piston-driven shock is generated by the fast and/or impulsive expansion of a driver in all directions, like in an explosion. In our case, the fast and impulsive expansion of the HC drives the shock. During this driving phase, the shock wave propagation is controlled by the motion of the piston, and the stand-off distance between shock and piston increases. Once the piston decelerates, the shock detaches and can continue its propagation, although now without additional energy supply by the piston. The piston can be slower than the shock, and has to accelerate rapidly. \citet{Zic2008} found that the shock-formation time and shock-formation distance are approximately proportional to the acceleration phase duration of the piston, shorter for a higher source speed and acceleration. Therefore, the HC kinematics demonstrated in \autoref{fig:SHC_vel} and \autoref{fig:HC_exp_vel} can be used to interpret the short shock-formation time and low shock-formation height in our event.

\subsection{Thermal Properties of the Eruption}

We analyze the thermal properties of the HC and its piston-driven shock through the differential emission measure (DEM) method, which measures the contribution of plasma emission in a given temperature range. To reconstruct the DEM, we run the routine ``XRT\li dem\li interative2.pro" in the Solar Software (SSW) package for each pixel in the region of interest. The code was originally developed by \citet{Weber2004} and has been modified slightly to work with AIA data \citep{Cheng2012}. \autoref{fig:Temper} (a) and (b) present the EM in two different temperature ranges for the HC driven shock and the HC. One can see that the shock region is primarily dominated by the plasma with $\log T = 6.1- 6.5$, and the HC is heated to a temperature of $\log T = 6.8-7.2$. In \autoref{fig:Temper} (c) we plot the sum of the DEM curve of each pixel in the region shown in \autoref{fig:Temper} (a) and (b). Apparently, there are two peaks in the figure. The major contribution to the first peak at $\log T = 6.2$ is primarily from the shock region. There may be also some contribution from the plasma in the surrounding regions. The second peak at $\log T = 6.9$ corresponds to the HC structure. The thermal properties of the HC and its driven shock in our event are consistent with results obtained previously for large-scale events \citep[e.g.,][]{Cheng2014,Su2015}.

\subsection{CME in LASCO Field of View}

Due to the high speed of the HC, it propagates to the LASCO/C2 FOV although suffering a significant deceleration. \autoref{fig:cme} (a) and (b) show two snapshots during its propagation. Red dashed lines mark the leading edge of the CME. Its internal structure is quite diffuse and does not show the classical three components: core, cavity, and front. Probably the material in mini-filaments involved in the eruption fell back to the surface so that they could not form a core in the LASCO FOV. We also notice that the propagation direction changes from southwest in the AIA FOV to equatorward in the LASCO FOV. The weak track, appearing in the distance-time plot along slice 3 in \autoref{fig:SHC_vel} (a), also can reveal the deflection of the HC at the edge of the AIA FOV. It is presumably influenced by the overlying loops above the HC whose footpoints as marked by the red star in \autoref{fig:cme} (c). The magnetic field lines in \autoref{fig:cme} (c) are derived from the Potential Field Source Surface (PFSS) method \citep{Schrijver2003}.
The characteristics of the CME are presented in \autoref{fig:cme_vel}. A linear fit to the height-time data in \autoref{fig:cme_vel} (a) yields an averaged velocity of 390 km s$^{-1}$. To derive the CME mass, we assume that it propagates in the plane of the sky and calculate the electron column density and the resultant mass in each pixel according to the Thomson scattering theory \citep{Feng2015a,Feng2015b}. The total mass is then derived by summing the mass in the area defined by the red dashed lines in \autoref{fig:cme}. We find that the CME mass increases with time and later reaches an almost constant value. It has a magnitude of $10^{14}$ g, but is subject to a large uncertainty in the background subtraction. Because of the existence of a preceding CME indicated by the white arrow in \autoref{fig:cme} (a), the mass calculated from base-difference images may be under-estimated. Therefore, we subtract two different backgrounds before mass calculation. One is a 12-hour minimum image calculated as the minimum values of images from 00:45 UT to 12:57 UT, the other is a pre-event image at 03:45:09~UT. In \autoref{fig:cme_vel} (b), the corresponding mass is delineated by the dashed line in the former case, by the dash dotted line in the later case. Their average is indicated by the solid line. Based on the computed mass, height, and velocity, we estimate the potential and kinetic energy of the CME. The obtained results are illustrated in \autoref{fig:cme_vel} (c) and (d). The magnitude of the potential energy is on an order of $10^{30}$ erg, which is about one order of magnitude larger than the kinetic energy.

\section{Discussions and Conclusions}

We have made comprehensive study on the small-scale short-duration eruption associated with mini-filaments occurred on November 4th, 2015. The source region of the eruption is a small-compact area with multiple magnetic polarities located at the edge of NOAA AR~12445 close to the west solar limb. \citet{Kumari2017} analyzed the same shock event. In our paper we focused not only on the analysis of the type II burst and shock, but also on the analysis of its driver, i.e., the CME. We did not use the shock to represent the near-Sun kinematics of the CME as in \citet{Kumari2017}. We present detailed analyses of the CME hot channel itself and its temporary and spatial relationship with the shock. Through the study of HC-shock relationship, we then can infer why the shock has an unusual high starting frequency and low formation height which was not included in \citet{Kumari2017}. The unusual large acceleration and high speed of the HC are the cause of high starting frequency. Concerning the type II dynamic spectra, we used higher-resolution data in which band split can be observed which allows us to further infer important parameters related to the shock than \citet{Kumari2017}. Therefore, the main conclusions are listed below:
\begin{itemize}
\item[-] Although the kinematics of the HC is in general consistent with the scenario of three-phase evolution, it has a very short duration (less than 2 minutes) in the main acceleration phase, and has an exceptional high maximal acceleration rate ($\sim$50 km s$^{-2}$), peak velocity ($\sim$1800 km s$^{-1}$), and maximal deceleration rate ($\sim$30 km s$^{-2}$), comparing to the kinematic statistics in \citet{Zhang2006}. We also detect a very fast expansion along the HC flank with a peak velocity of about $\sim1350\,\mathrm{km\,s^{-1}}$.
\item[-] The fast motion of the HC along different directions acts as a piston and drives a fast shock. The kinematic measurements including velocity, acceleration, stand-off distance, etc., of the HC and its driven shock fit the piston-driven shock scenario. The analyses of the associated type II burst together with the AIA imaging data reveals very low starting height of the type II burst, at about $1.1~\mathrm{R_{\odot}}$. The type II band split is used to constrain the decrease of the shock compression ratio (2.2 to 1.3), Alfv{\'e}nic mach number (1.9 to 1.3), from where the Alfv{\'e}n speed (1000 to 400~$\mathrm{km\,s^{-1}}$) and the coronal magnetic field strength (13 to 0.5 G) are derived in the heliocentric distance from 1.1 to 2.3~$\mathrm{R_{\odot}}$.
\item[-] The thermal properties of the HC and its driven shock are consistent with those derived for large-scale events based on the DEM analyses.
\item[-] The CME observed in the LASCO/C2 FOV has a small leading-edge velocity ($390\,\mathrm{km\,s^{-1}}$), small mass ($\sim10^{14}\,\mathrm{g}$), potential ($\sim10^{30}\,\mathrm{erg}$) and kinetic ($\sim10^{29}\,\mathrm{erg}$) energy.
\end{itemize}

Concerning flare and CME energetics, \citet{Emslie2012} and \citet{Feng2013} have found the energy equal-partition between the flare and the CME for large-scale eruptions. For this small-scale event, the total radiated output classified as the `final' energy \citep{Emslie2005} is used to estimate the flare energy, and the sum of the potential and kinetic energy is a measure of the CME energy. The GOES and RHESSI observations show impulsive and short duration M1.9 flare, shorter than that of those large-scale eruptions which often last for tens of minutes to several hours in previous studies \citep{Chamberlin2012,Feng2013,Cheng2014,Cheng2015}. \citet{Chamberlin2012} found that the total radiated output of flares depends more on the flare duration than the typical GOES X-ray peak magnitude classification. Here, we estimate the output irradiance by using the data of the EUV Variability Experiment \citep[EVE;][]{Woods2012} onboard SDO. The light curves of the 0.1-7 nm waveband derived from the Extreme Ultraviolet Spectro-Photometer (ESP) is shown in \autoref{fig:eve}, and then used to quantify the radiative output of the EUV emission. The integration of the background-subtracted irradiance over the flare duration yields the radiated output in 0.1-7 nm is about $1.6\times10^{30}$~erg. The total radiated output is a few times larger than this value \citep{Emslie2012}. We can find that the flare probably consumes similar amount of magnetic energy to the CME ($4\times10^{30}$~erg) in this event. Due to the short duration of the flare, compared with the large-scale eruptions \citep{Emslie2012,Feng2013}, the released magnetic energy of this flare is about one order of magnitude lower. \citet{Aschwanden2016} selected 399 M and X class flare events and obtained the parameters of the associated CMEs. We find that the CME mass and energy, in this event, lies in the lower end of their statistics. Although the absolute quantities of the flare and CME energy are smaller, the very similar partition of the flare and CME energy along with a similar multi-phase kinematics may imply that small- and large-scale events share similar relationship between flares and CMEs.
Although we have found some common characteristics between the small- and large-scale eruptions in terms of the flare-and-CME temporal and energetic relationships, we also need to understand what makes this event kinematically special, e.g., a very short impulsive acceleration phase, and a very high acceleration. We explore the Erupting Flux Rope model proposed by \citet{Chen1989}, in which there exists a relationship between the acceleration $a$ of an MFR (or a HC, its manifestation in AIA) and its geometrical size:
 \begin{equation}
 a = \frac{d^2Z}{dt^2} \propto [R\ln(8R/a_f)]^{-2},
\end{equation}
\begin{equation}
 R = \frac{Z^2+S_{f}^2/4}{2Z}.
\end{equation}
where $Z$ is the height of the centroid at the FR apex; $R$ is the major radius of the current channel assumed to be uniform along the MFR; $a_f$ is the minor radius at the footpoint which is also assumed as an invariant; $S_f$ is the distance of FR footpoint separation. \citet{Chen2003} found a scaling law related to the acceleration of CMEs and $S_{f}$: there are two critical heights scaled with $S_{f}$ in condition that the studied structure is an MFR, and the scale is given by $Z_{\ast} = S_f /2$ and $Z_m \backsimeq 1.5 S_f $, such that the height $Z_{max}$ where the acceleration of the centroid of the MFR apex is maximal satisfies $Z_{\ast} < Z_{max}< Z_m$. We speculate that $S_f$ is very small and eventually leads to a very low $Z_{max}$. According to Equation (6) and (7), there is also a tendency that the smaller the major radius R is, the larger the acceleration of the MFR will be. Therefore, the large acceleration together with the low height at which the MFR reaches its maximal acceleration jointly produce a short duration of the impulsive acceleration phase.
\acknowledgements 
We are very grateful to  anonymous referee for very constructive comments and suggestions. We thank James Chen for his suggestions on the EFR model. We are also thankful to the World Data Centre of the Australian Bureau of Meteorology, Space Weather Services for the Culgoora and Learmonth radio spectrograph data.  SDO is a mission of NASA's Living With a Star Program, SOHO is a mission of international cooperation between ESA and NASA. This work is supported by the NSFC grants (11522328, 11473070, 11427803, U1731241) and by CAS Strategic Pioneer Program on Space Science, Grant No. XDA15010600, XDA15052200, XDA15320103, XDA 15320301. L.F. also acknowledges the specialized research fund from the State Key Laboratory of Space Weather for financial support. Y.S. is supported by the One Hundred Talent Program of CAS and NSFC 11473071. J.Z. is supported by US NSF AGS-1249270 and NSF AGS-1156120.
\bibliography{bibliography2}
\end{document}